\documentclass[11pt]{article}
\usepackage{jcappub}
\usepackage{amsmath}
\usepackage{amssymb}
\usepackage{accents}
\usepackage{subfigure}
\newcommand{\be}{\begin{equation}}
\newcommand{\ee}{\end{equation}}
\newcommand{\bea}{\begin{eqnarray}}
\newcommand{\eea}{\end{eqnarray}}
\newcommand{\bp}{\begin{pmatrix}}
\newcommand{\ep}{\end{pmatrix}}
\newcommand{\mh}{\mathcal{H}}
\newcommand{\mr}{\mathcal{R}}

\newcommand{\lsp}{\left(}
\newcommand{\rsp}{\right)}
\newcommand{\lsb}{\left[}
\newcommand{\rsb}{\right]}
\newcommand{\lsc}{\left\{}
\newcommand{\rsc}{\right\}}
\newcommand{\nn}{\nonumber}
\newcommand{\vp}{\varphi}
\newcommand{\vpb}{\vp_0}
\newcommand{\dvp}{\delta\vp}
\newcommand{\tdvp}{\tilde\dvp}
\newcommand{\tp}{\tilde\phi}
\newcommand{\tb}{\tilde B}
\newcommand{\te}{\tilde E}
\newcommand{\ba}{\begin{appendix}}
\newcommand{\ea}{\end{appendix}}

\title{Cosmological perturbation of Unimodular Gravity and General Relativity are identical}

\author{Abhishek Basak,}
\author{Oph\'elia Fabre}
\author{and S. Shankaranarayanan}

\affiliation{School of Physics, Indian Institute of Science Education and
Research Thiruvananthapuram (IISER-TVM), Trivandrum 695016, India}

\emailAdd{abhishek@iisertvm.ac.in}
\emailAdd{ophelia.fabre@iisertvm.ac.in}
\emailAdd{shanki@iisertvm.ac.in}

\abstract{ 
Unimodular Gravity (UG) is a restricted version of General Relativity (GR) in which the determinant of the metric is a fixed function and the field equations are given by the trace-free part of the full Einstein equations. The background equations in UG and GR are identical. It was recently claimed that, the first order contribution in the temperature fluctuation of the Cosmic Microwave Background (CMB) in UG is different from GR. In this work, we calculate the first order perturbation equations in UG and show that the Sachs-Wolfe effect in UG, in terms of gauge invariant variables, is identical to GR. We also show that the second order perturbation equation of Mukhnanov-Sasaki variable in UG, is identical to GR. The only difference comes from the gauge choices due the constraints on the metric determinant. Hence, UG and GR are identical and indistinguishable in CMB data on large scales.}
 
\begin{document}

\maketitle

\section{Introduction}

The cosmological constant problem in cosmology is an unsolved
issue. In the theory of General Relativity (GR), the cosmological
constant is put in the Einstein-Hilbert action by hand, and is related
to the vacuum energy density of all the matter fields.
From observations, like Cosmic Microwave Background data \cite{Ade:2013zuv,planck2015planck} and Supernovae surveys \cite{riess1998observational}, we know that the contribution of the cosmological constant in the total energy density of the universe is non-zero. The problem is that the
theoretical value of the cosmological constant is 60-120 orders of
magnitude higher than the observed value \cite{Weinberg:1989ab, Padmanabhan:2002ji}. This
leads to a very large discrepancy between theory and observation.
There have been several proposals in the literature
\cite{Weinberg:1989ab, martin2012everything} to solve this discrepancy. Out of these, trace-free gravity does
not involve any new physics and is a particular case of General
Relativity. The trace-free part of the Einstein equation naturally
gives the cosmological constant which is no longer related to the
vacuum energy density. Hence, the value of the cosmological constant
can be fixed from observation.

The trace-free part of Einstein equation can be obtained in the
context of Unimodular Gravity (UG) \cite{Anderson:1971pn,
  Unruh:1988in}. In UG, the determinant of the metric ($\sqrt{-g}$) is
a fixed function\cite{Finkelstein:2000pg}. Hence, the variation of
$\sqrt{-g}$ with respect to the metric is zero. The variation of
Einstein-Hilbert action, 
\be S_{E-H}= \int
d^{4}x\sqrt{-g}\lsb\frac{M^{2}_{P}}{2}R + \mathcal{L}_{m}\rsb, 
\label{eh:action}
\ee
subjected to the constraint on $\sqrt{-g}$, leads to the trace-free
part of the Einstein equation, \be \hat{G}_{\mu \nu}=8\pi
G\hat{T}_{\mu \nu}.
\label{tf:ee}
\ee 
In the Einstein-Hilbert action, $M_{P}^2=\frac{1}{8\pi G}$ is the
reduced Planck mass, $R$ is the Ricci scalar and $\mathcal{L}_{m}$ is
the matter counterpart of the action. $\hat{G}_{\mu \nu}$ and
$\hat{T}_{\mu \nu}$ are the trace-free part of the Einstein tensor and
energy-momentum tensor respectively, 
\be \hat{G}_{\mu \nu} = R_{\mu
  \nu}-\frac{1}{4}g_{\mu \nu}R, \qquad \hat{T}_{\mu \nu} = T_{\mu
  \nu}-\frac{1}{4}g_{\mu \nu}T.\label{tf:et}, 
\ee 
where $T$ is the trace of the full energy-momentum tensor
$T_{\mu\nu}$.

In Refs.~\cite{Ellis:2010uc, Ellis:2013uxa}, it has been shown that,
in UG, the information of potential enters the acceleration equation
through the energy conservation equation. Hence, like in GR, one can
study inflationary scenario in UG. Several authors have studied cosmological implications of UG
theory \cite{Jain:2012gc, Cho:2014taa}. In Refs. \cite{Shaposhnikov:2008xb, Blas:2011ac,
  GarciaBellido:2011de} the authors have applied UG theory in inflation
and dark energy models.

In FLRW background \cite{Ellis:2010uc, Ellis:2013uxa},
there are no differences in the equations of motion in UG and
GR. Quantum theory of UG has been investigate by several authors \cite{Smolin:2009ti, Smolin:2010iq}. In Refs. \cite{Eichhorn:2013xr, Bufalo:2015wda} it has been shown that the quantum theory of UG and GR can be different.

However, in Ref.~\cite{Gao:2014nia}, the authors claimed that,
the expression of first order contribution in the temperature fluctuation of the Cosmic Microwave Background (CMB) in UG is different 
from GR. The additional term in UG would be a dipole term whose amplitude is expected to be
small. In this work we explicitly show that the expression of first order perturbation counterpart in CMB
temperature fluctuations in UG is identical to that in GR. Furthermore,
we explicitly calculate the second order perturbation in UG and show
that the second order equations of motion are identical in UG and GR.

In Section~\ref{background_equations}, we introduce the background
equations in the UG context. Then, in Section~\ref{perturbed_metric}, we present the perturbed metric and the unimodular constraint. 
In Section~\ref{first_order}, we express the first order perturbations and define the gauge invariant quantities,
before focusing on the implications of UG on the Sachs-Wolfe contribution in the CMB temperature anisotropies in Section~\ref{SW_CMB}. 
The study is then extended in Section~\ref{second_order} to second order perturbations. The equation of motion obtained
for first order Mukhanov-Sasaki variable is the same as in
GR, whereas the equation of motion obtained with second order
Mukhanov-Sasaki variable is also same as GR for particular gauge
choices. As a consequence, in the light of perturbation theories, UG
and GR are same. Finally, in Section~\ref{conclusion}, we discuss the results about the differences
between GR and UG in the light of the problems due to gauge choices and
conclude.

%%%%%%%%%%%%%%%%%%%%%%%%%%%%%%%%%%%%%%%%%%%%%%%%%%%%%%%%%%%%%%%%%%%%%%%%%%%%%%%%%%%%%%%%%%%%%%
\section{Background equations}\label{background_equations}

In UG, $\sqrt{-g}$ is a fixed function. As a consequence, one can take FLRW metric as a
background metric. Then the fixed function associated with the
determinant of the metric is a function of the scale factor
($a(t)$). In this Section, we consider the FLRW metric described by the
following line element in the cosmic time ($t$) and position
($\vec{x}$): 
\be ds^{2}=dt^{2}-a^{2}\lsp t\rsp d\vec{x}^{2}.
\label{frw}
\ee
In the following, we give the components of the background trace-free Einstein equation and compare the results with GR.
%%%%%&&&&&&&&&&&&&&&&&&&&&
\subsection{Components of trace-free Einstein tensor}
The background components of the trace-free Einstein tensor (\ref{tf:et}) are
\be
\hat{G}_{t t}=\frac{3}{2}\lsp H^{2}-\frac{\ddot{a}}{a}\rsp, \qquad
\hat{G}_{i j}=\frac{1}{2} a^{2} \lsp H^{2}-\frac{\ddot{a}}{a}\rsp \delta_{i j},
\label{tf:et1}
\ee
where $H=\dot{a}/a$ is the Hubble parameter and the `\emph{dot}' denotes the derivative with respect to time. `$i$' is the spatial index which represents the spatial coordinates ($x$, $y$, $z$).
%%%%%&&&&&&&&&&&&&&&&&&&&&
\subsection{Components of trace-free energy-momentum tensor}
The expression of the energy-momentum tensor for a scalar field ($\vp$) is given by
\be
T_{\mu \nu}=\partial_{\mu}\vp\partial_{\nu}\vp-g_{\mu \nu}\lsp\frac{1}{2}\partial_{\sigma}\vp\partial^{\sigma}\vp-V\lsp\vp\rsp\rsp,
\label{em}
\ee
where $\partial_{\mu}$ is the partial derivative with respect to $x^{\mu}$ and $V\lsp\vp\rsp$ is the potential. The components of the energy-momentum tensor are
\be
T_{t t}=\frac{1}{2}\dot{\vp}^{2}+V\lsp\vp\rsp, \qquad
T_{i j}=a^{2}\lsp\frac{1}{2}\dot{\vp}^{2}-V\lsp\vp\rsp\rsp\delta_{i j}.
\label{em:1}
\ee
Subsequently, the energy density and pressure of the scalar field are
\be
\rho=T^{t}_{t}=\frac{1}{2}\dot{\vp}^{2}+V\lsp\vp\rsp, \qquad
p=-T^{i}_{i}=\frac{1}{2}\dot{\vp}^{2}-V\lsp\vp\rsp.
\label{em:2}
\ee
One can write the trace-free components of the energy-momentum tensor as
\be
\hat{T}_{t t}=\frac{3}{4}\dot{\vp}^{2}, \qquad
\hat{T}_{i j}=\lsp\frac{1}{4}a^{2}\dot{\vp}^{2}\rsp\delta_{i j}.
\label{tf:em1}
\ee
%%%%%&&&&&&&&&&&&&&&&&&&&&&&&&&&&&&&&&&
\subsection{Continuity equation}
As, the whole formalism is generally covariant, the usual conservation equation of the energy-momentum tensor is satisfied \cite{Weinberg:1989ab}. The conservation equation of the scalar field energy-momentum tensor is given as 
\be
T^{\mu}_{\nu;\mu}=0,
\label{cont}
\ee
where `$;$' denotes the covariant derivative. Using (\ref{em:2}) for $\nu=0$, Eqn. (\ref{cont}) gives the energy conservation equation
\be
\dot{\rho}+3H\lsp\rho+p\rsp=0,
\label{cont1}
\ee
which in terms of the scalar field ($\vp$) gives the Klein-Gordon equation for the scalar field
\be
\ddot{\vp}+3H\dot{\vp}+V_{,\vp}=0,
\label{kg}
\ee
where ($_{,\vp}$) denotes partial derivative with respect to $\vp$.

%%%%%&&&&&&&&&&&&&&&&&&&&&&&&&&&&&&&&&&
\subsection{Components of trace-free Einstein equation}

Using (\ref{tf:et1}) and (\ref{tf:em1}), the non-zero components of trace-free Einstein equation (\ref{tf:ee}) can be written as
\be
\frac{\ddot{a}}{a}-H^{2}=-4\pi G \dot{\vp}^{2}.
\label{tf:ee1}
\ee
Here it should be noted that unlike the full Einstein equation, the trace-free Einstein equation gives us the same equation for the ($t,t$) and ($i,j$) components as (\ref{tf:ee1}).
Apparently it looks like the potential does not enter the acceleration equation in case of trace-free Einstein equation. However, after multiplying both sides of (\ref{tf:ee1}) with the factor ($6H$) and using (\ref{kg}), one gets the Friedmann equation by integrating
\be
H^{2}=\frac{8\pi G}{3}\lsp\frac{1}{2}\dot{\vp}^{2}+V\lsp\vp\rsp\rsp+C,
\label{tf:ee2}
\ee
where $C$ is the integration constant. 
Substituting $H^{2}$ from (\ref{tf:ee2}) in (\ref{tf:ee1}), the acceleration equation can be written as 
\be
\frac{\ddot{a}}{a}=-\frac{8\pi G}{3}\dot{\vp}^{2}+\frac{8\pi G}{3}V\lsp\vp\rsp+C.
\label{tf:ee3}
\ee
Note that, Friedmann equation (\ref{tf:ee2}) and acceleration equation (\ref{tf:ee3}) look like the equations obtained from the full Einstein equations in GR. This is not surprising as one can write the trace-free Einstein equation (\ref{tf:ee}) in a different way 
\be
R_{\mu \nu}-\frac{1}{2}g_{\mu \nu}R+g_{\mu \nu}\hat{\Lambda}=8\pi G T_{\mu \nu},
\label{tf:eea}
\ee
where $\hat{\Lambda}=\frac{1}{4}\lsp R+8\pi GT\rsp$.
For FLRW metric, the expression of $R$ and $T$ can be obtained as $R=-6\lsp H^{2}+\frac{\ddot{a}}{a}\rsp$ and $T=-\dot{\vp}^{2}+4V\lsp\vp\rsp$ respectively. Finally using (\ref{tf:ee1}) and (\ref{tf:ee2}), from the expression of $\hat\Lambda$ it can be verified that $\hat{\Lambda}=-3C$. This exercise can be used as a proof that $\hat{\Lambda}$ in (\ref{tf:eea}) is a constant in time. It is to be noted that, $\hat\Lambda$ is not a part of the action (\ref{eh:action}) and $\hat\Lambda$ does not depend upon the vacuum fluctuations \cite{Weinberg:1989ab}. 
%%%%%&&&&&&&&&&&&&&&&

%%%%%%%%%%%%%%%%%%%%%%%%%%%%%%%%%%%%%%%%%%%%%%%%%%%%%%%%%%%%%%%%%%%%%%%%%%%%%%%%%%%%%%%%%%%%%%
\section{Perturbed metric and unimodular constraint}\label{perturbed_metric}

In our study, we will only work with the scalar part of the metric perturbations. From this section onwards, we specify background quantities with subscript ($0$); subscripts ($1$) and ($2$) for first order and second order perturbations respectively. In conformal time $\eta$ ($d\eta=dt/a$), the metric components up to second order can be written as 
\bea
\nn g_{\eta \eta} &=& a^{2}\lsp 1 + 2\phi_1 + \phi_2 \rsp,\\
\nn g_{\eta i} &=& - a^{2}\lsp \partial_{i}{B_1} + \frac{1}{2}\partial_{i}{B_2}\rsp, \\
    g_{i j} &=& - a^{2}\lsp \delta_{i j} + 2 C_{1 i j} + C_{2 i j}\rsp.
\label{me:cov}
\eea
Consequently, the contravariant form becomes,
\bea
\nn g^{\eta \eta} &=& a^{-2} \lsc 1 - 2\phi_1 - \lsp \phi_2 - 4\phi_1^{2} + \partial_{i}{B_1}\partial_{i}{B_1}\rsp\rsc,\\
\nn g^{\eta i} &=& -a^{-2} \lsc \partial_{i}{B_1} + \lsp\frac{1}{2}\partial_{i}{B_2} - 2\phi_1\partial_{i}{B_1} - 2\partial_{k}{B_1}C_{1 k i}\rsp\rsc, \\
g^{i j} &=& -a^{-2} \lsc\delta_{i j} - 2 C_{1 i j} - \lsp C_{2 i j} - 4 C_{1 i k} C_{1 k j} + \partial_{i}{B_1}\partial_{j}{B_1}\rsp\rsc.
\label{me:cov}
\eea
In the above expression $C_{1ij}$ and $C_{2ij}$ are given by
\bea
\nn C_{1ij} &=& - \psi_1 \delta_{i j} + \partial_{i j}{E_1}, \\
    C_{2ij} &=& - \psi_2 \delta_{i j} + \partial_{i j}{E_2},
\label{me:addl}
\eea

\noindent where $\lsp\phi_1, \psi_1, B_1, E_1\rsp$ are first order quantities and  $\lsp\phi_2, \psi_2, B_2, E_2\rsp$ are second order quantities. 

A metric $\lsp g_{\mu \nu}\rsp$ can be written in terms of background metric $\lsp\eta_{\mu \nu}\rsp$ and perturbation $\lsp h_{\mu \nu}\rsp$ as, 
\be
g_{\mu \nu} = \eta_{\mu \nu} + h_{\mu \nu}. 
\label{me:gen}
\ee
Therefore, the determinant of the metric $g_{\mu \nu}$, up to second order, can be written as, 
\be
\sqrt{-g} = \sqrt{-\eta}\lsb 1 + \frac{1}{2}\eta^{\mu \nu}h_{\mu \nu} -\frac{1}{4}h^{\mu \nu}h_{\mu \nu} + \frac{1}{8}\lsp \eta^{\mu \nu} h_{\mu \nu} \rsp^{2} + \mathcal{O}\lsp h_{\mu \nu}^{3}\rsp\rsb.
\label{me:det}
\ee
As a consequence, the unimodular constraint, $\delta\lsp\sqrt{-g}\rsp=0$, gives the following constraints on the first and second order metric quantities respectively,
\bea
    &&\phi_1 - 3\psi_1 + \Delta E_1 = 0,\label{ugc:f}\\
\nn &&\frac{1}{2}\lsp\phi_2 - 3\psi_2 + \Delta E_2\rsp  + \frac{3}{2}\phi_1^{2} - \frac{1}{2}\lsp\partial_{i}{B_1}\rsp^{2} + \frac{15}{2}\lsp\psi_1\rsp^{2} + \lsp\partial_{i j}E_1\rsp^{2} - 3\phi_1\psi_1 + \Delta E_1\lsp\phi_1- 5\psi_1\rsp \\ 
    &&+ \frac{1}{2}\lsp\Delta E_1\rsp^{2} = 0.\label{ugc:s}
\eea

The infinitesimal coordinate transformation (or gauge transformation), up to second order, is given by \cite{Mukhanov:1996ak},
\be
x^{\mu}\rightarrow\tilde{x}^{\mu}=x^{\mu}+\xi^{\mu},
%x^{\mu}\rightarrow\tilde{x}^{\mu}=x^{\mu}+\xi^{\mu}+\frac{1}{2}\xi^{\mu}_{,\nu}\xi^{\nu},
\label{ct:gen}
\ee
where the gauge generator $\xi^{\mu}$, up to second order, is $\xi^{\mu}=\xi^{\mu}_{1}+\frac{1}{2}\xi^{\mu}_{2}$. The components of $\xi^{\mu}$ in first order and second order are given as $\xi^{\mu}_{1}=\lsc\alpha_{1}, \lsp\beta_{1 ,i}\rsp\rsc$ and $\xi^{\mu}_{2}=\lsc\alpha_{2}, \lsp\beta_{2 ,i}\rsp\rsc$, respectively. The construction of gauge invariant variables corresponds to the elimination of gauge generators, which in this case are $\alpha_{1}, \beta_{1}, \alpha_2$ and $\beta_{2}$.

Under the above coordinate transformation, an arbitrary tensor \textbf{Q} transforms as 
\be
\textbf{Q}\rightarrow\tilde{\textbf{Q}}=\lsp 1-\mathcal{L}_{\xi}+\frac{1}{2}\mathcal{L}_{\xi}\mathcal{L}_{\xi}\rsp \textbf{Q},
\label{tt:gen}
\ee
where $\mathcal{L}_{\xi}$, is the Lie derivative with respect to $\xi^{\mu}$. Up to second order, $\textbf{Q}$ can be expressed as 
\be
\textbf{Q} = \textbf{Q}_{0}  + \delta\textbf{Q}_{1} + \frac{1}{2}\delta\textbf{Q}_{2}.
\ee
Therefore, transformations of the background, first order and second order of $\textbf{Q}$ are given as 
\bea
\nn \tilde{\textbf{Q}}_{0}&=&\textbf{Q}_{0},\\
\nn \tilde{\delta\textbf{Q}}_{1}&=&\delta\textbf{Q}_{1}-\mathcal{L}_{\xi_{1}}\textbf{Q}_{0},\\
    \tilde{\delta\textbf{Q}}_{2}&=&\delta\textbf{Q}_{2}-\mathcal{L}_{\xi_{2}}\textbf{Q}_{0}+
    \mathcal{L}_{\xi_{1}}\mathcal{L}_{\xi_{1}}\textbf{Q}_{0}-2\mathcal{L}_{\xi_{1}}\delta\textbf{Q}_{1}.
    \label{tt:gen1}
\eea
To derive the above expressions, we have used, $\mathcal{L}_{\xi}=\lsp\mathcal{L}_{\xi_{1}}+\frac{1}{2}\mathcal{L}_{\xi_{2}}\rsp$.
For scalar $\lsp S\rsp$, vector $\lsp V_{\mu}\rsp$ and tensor $\lsp t_{\mu \nu}\rsp$, the Lie derivatives, on all order, are given as 
\bea
\nn \mathcal{L}_{\xi}S &=& \xi^{\lambda}S_{,\lambda},\\
\nn \mathcal{L}_{\xi}V_{\mu} &=& \xi^{\lambda}V_{\mu,\lambda}+\xi^{\lambda}_{,\mu}V_{\lambda},\\
    \mathcal{L}_{\xi}t_{\mu \nu} &=& \xi^{\lambda}t_{\mu \nu,\lambda}+\xi^{\lambda}_{,\nu}t_{\mu    \lambda}+\xi^{\lambda}_{,\mu}t_{\nu \lambda}.
    \label{lie:svt}
\eea
The above Lie derivatives are used to obtain the transformations of scalar quantities (field fluctuation, $\dvp$ and fluctuation of the energy density, $\delta\rho$) and the components of metric fluctuation, $g_{\mu \nu}$.

%%%%%%%%%%%&&&&&&&&&&&&&&&&&&&&&&&&&&&&&&&&&&&&&&&&&&&&&&
\section{First order perturbations}\label{first_order}

In this Section, we derive the gauge invariant quantities for UG in first order perturbation. 

Using the perturbed field equations for UG, we derive the equation of motion for the gauge invariant quantities. The equations of motion of the gauge invariant metric perturbations, $\Phi_1$ and $\Psi_1$, have already been studied by various authors. We then compare UG with GR based on first order Mukhanov-Sasaki variable and curvature perturbation on uniform density hypersurfaces. We show that the equation of motion for the gauge invariant quantities are identical in UG and GR. 

We then look into observational signature of UG in the Cosmic Microwave Background (CMB). Recently, it has been claimed that the expression of the temperature anisotropies of the CMB, with regard to the Sachs-Wolfe contribution, are different in UG and GR \cite{Gao:2014nia}. This discrepancy in the CMB temperature anisotropies is supposed to appear as a dipole term. In contrast, we show that, in UG, the relation between the first order CMB temperature fluctuation and gauge invariant metric perturbation is identical to GR. Therefore, CMB observations can not distinguish between UG and GR.  

In UG, first order perturbed field equations can be obtained from  
\be
\delta\hat{G}^{\mu}_{\nu \lsp 1\rsp}=8\pi G \delta\hat{T}^{\mu}_{\nu \lsp 1\rsp},
\label{tf:eep}
\ee
where $\delta\hat{G}^{\mu}_{\nu \lsp 1\rsp}$ and $\delta\hat{T}^{\mu}_{\nu \lsp 1\rsp}$ are the trace-free part of the perturbed Einstein tensor and energy-momentum tensor respectively.
\subsection{Perturbed trace-free energy-momentum and Einstein tensor}
\noindent\underline{\emph{Components of perturbed trace-free energy-momentum tensor}} --- 
\bea
\nn\delta\hat{T}^{\eta}_{\eta \lsp 1\rsp}&=&-\frac{3}{2a^{2}}\lsb\vpb^{\prime}\lsp\phi_1\vpb^{\prime}-\delta\vp^{\prime}_1\rsp\rsb,\\
\nn\delta\hat{T}^{i}_{i \lsp 1\rsp}&=&\frac{1}{2a^{2}}\lsb\vpb^{\prime}\lsp\phi_1\vpb^{\prime}-\delta\vp^{\prime}_1\rsp\rsb,\\
\nn\delta\hat{T}^{\eta}_{i \lsp 1\rsp}&=&\frac{1}{a^{2}}\lsp\vpb^{\prime}\delta\vp_1\rsp_{,i},\\
\delta\hat{T}^{i}_{j \lsp 1\rsp}&=& 0 \qquad \lsp i\neq j\rsp,
\label{tf:emp}
\eea
where $\vpb$ is the background scalar field.
%%%%%%%%%%%%&&&&&&&&&&&&&&&&&&&&&&&&&&&&&&&&&&&&&&&&&&&&
\\

\noindent\underline{\emph{Components of perturbed trace-free Einstein tensor}} ---
\bea
\nn\delta\hat{G}^{\eta}_{\eta \lsp 1\rsp}&=&-\frac{1}{2a^{2}}\lsb 6\lsp\mh^{2} - \mh^{\prime}\rsp\phi_1 + 3\mh\psi^{\prime}_1 - 3\psi^{\prime \prime}_1 - 3\mh\phi^{\prime}_1 - \Delta A\rsb,\\
\nn\delta\hat{G}^{i}_{i \lsp 1\rsp}&=&\frac{1}{2a^{2}}\lsb 2\lsp\mh^{2} - \mh^{\prime}\rsp\phi_1 + \mh\psi^{\prime}_1 - \psi^{\prime \prime}_1 - \mh\phi^{\prime}_1 - \Delta B + 2\partial_{i i}D\rsb,\\
\nn\delta\hat{G}^{\eta}_{i \lsp 1\rsp}&=&\frac{2}{a^{2}}\lsp\psi^{\prime}_1 + \mh\phi_1\rsp_{,i},\\
\delta\hat{G}^{i}_{j \lsp 1\rsp}&=&\frac{1}{a^{2}}\partial_{i}\partial_{j}D \qquad \lsp i\neq j\rsp,
\label{tf:etp}
\eea
where $A = \phi_1 + 2\psi_1 - \mh\lsp B_1-E_1^{\prime}\rsp + B_1^{\prime} - E_1^{\prime\prime}$, 
$B = \phi_1 + \mh\lsp B_1 - E_1^{\prime}\rsp + B_1^{\prime} - E_1^{\prime\prime}$, $D = \phi_1 - \psi_1 + 2\mh\lsp B_1 - E_1^{\prime}\rsp + B_1^{\prime} - E_1^{\prime\prime}$ and $\mh=a^{\prime}/a$ is the Hubble parameter in conformal time.
%%%%%%%%%&&&&&&&&&&&&&&&&&&&&&&&&&&&&&&&&&&&&&&&&&&&&&&&
\\

\noindent\underline{\emph{Components of perturbed trace-free Einstein equation}} ---

From the off-diagonal ($i, j$) components of perturbed trace-free Einstein equation, in UG one gets 
\be
D = \phi_1 - \psi_1 + 2\mh\lsp B_1 - E_1^{\prime}\rsp + B_1^{\prime} - E_1^{\prime\prime} = 0,
\label{tf:eep1}
\ee
which is identical to GR. From the ($\eta, i$) components of perturbed trace-free Einstein equation, in UG one gets 
\be
\psi^{\prime}_1 + \mh\phi_1 = 4\pi G\lsp\vpb^{\prime}\delta\vp_1\rsp,
\label{tf:eep2}
\ee
which is also identical to GR. On the contrary, in UG, the perturbed ($\eta, \eta$) component of trace-free Einstein equation gives 
\be
\psi_1^{\prime \prime} + \mh \lsp\phi_1^{\prime} - \psi_1^{\prime}\rsp + \frac{\Delta}{3} A = 8\pi G \vpb^{\prime}\delta\vp^{\prime}_1,
\label{tf:eep3}
\ee
which is different compared to GR. To get the above equation we have used the following background equations,
\bea
\nn &&\mh^{2}-\mh^{\prime}=4\pi G \vpb^{\prime 2},\\
&& \vpb^{\prime\prime} + 2\mh\vpb^{\prime} + a^{2}V_{,\vp} = 0,
\label{back:frd}
\eea
where ($_{,\vp}$) denotes the partial derivative with respect to the background field. 
In UG, ($i, i$) component of the perturbed trace-free Einstein equation, gives the same equation as (\ref{tf:eep3}) by summing over the `$i$' indices. Therefore, we do not write it separately. 

The perturbed Klein-Gordon equation in UG is
\be 
\dvp^{\prime\prime}_1 + 2\mh\dvp^{\prime}_1 - \Delta\dvp_1 + a^{2}V_{, \vp \vp}\dvp_1 - \vp^{\prime}_0\lsp \phi_1 + 3\psi_1\rsp^{\prime} + 2 a^{2} V_{, \vp}\phi_1 - \vpb^{\prime} \Delta\lsp B_1 - E_1^{\prime}\rsp = 0.
\label{KG:1}
\ee
One can note that, similar to the background case, perturbed Klein-Gordon equation is also same as in GR. 
\subsection{Gauge invariant quantities} 

Under the gauge transformations (\ref{tt:gen1}) and (\ref{lie:svt}), the perturbed energy density and scalar fields transform as 
\bea
\nn \tilde{\delta\rho_{1}}&=&\delta\rho_{1}-\rho^{\prime}_{0}\alpha_{1},\\
    \tilde{\delta\vp_{1}}&=&\delta\vp_{1}-\vp^{\prime}_{0}\alpha_{1}.
    \label{tt:rph}
\eea
In the above equations, $\rho_0$ denotes the background energy density.
The first order metric perturbations transform as
\bea
\nn \tilde{\psi_1}&=&\psi_1 + \mh\alpha_{1},\\
\nn \tilde{\phi_1}&=&\phi_1 - \mh\alpha_{1} - \alpha^{\prime}_{1},\\
\nn \tilde{B_1} &=& B_1 + \alpha_{1} - \beta_{1}^{\prime},\\
    \tilde{E_1}&=& E_1 - \beta_{1}.
    \label{tt:metric}
\eea
By substituting expressions of metric perturbations (\ref{tt:metric}) in Eq.(\ref{ugc:f}), one can construct the unimodular constraint on the quantities $\alpha_{1}$ and $\beta_{1}$ as,
\be
\Delta\beta_{1}+\alpha^{\prime}_{1}+4\mh\alpha_{1}=0.
\label{ugc:f1}
\ee
In the present case, because of the unimodular constraint, one can choose either only $\alpha_{1}$, or only $\beta_{1}$, freely. Once $\alpha_{1}$ or $\beta_{1}$ is fixed, the other quantity is determined by the unimodular constraint. Therefore, gauge conditions like longitudinal Newtonian gauge (\textit{i.e.} $\tilde{B_1}=0$ and $\tilde{E_1}=0$) or transverse synchronous gauge (where we set, $\tilde{\phi_{1}}=0$ and $\tilde{B_{1}}=0$) can not be considered in unimodular gravity \cite{Gao:2014nia}. This is one big difference between unimodular gravity and GR. However, there are some gauge invariant quantities which can be constructed by choosing $\alpha_{1}$ or $\beta_1$.

Based on the above expressions, one can construct various first order gauge invariant quantities. On uniform field hypersurfaces $\lsp\tilde{\delta\vp_{1}}=0\rsp$, the expression of $\alpha_{1}$ is
\be
\alpha_{1}=\frac{\delta\vp_{1}}{\vp^{\prime}_{0}}.
\ee
Substituting $\alpha_1$ in the expression of $\tilde{\psi_1}$ leads to comoving curvature perturbation ($\psi_1$ on uniform field hypersurfaces)
\be
\mathcal{R}_{1} = \tilde\psi_1|_{\tdvp_1=0} = \psi_1 + \frac{\mh}{\vp^{\prime}_{0}}\delta\vp_{1}.
\label{ug:r1}
\ee
Following a similar procedure, other gauge invariant quantities can be constructed. For example, field fluctuation on uniform curvature $\lsp\tilde{\psi_1}=0\rsp$ hypersurfaces (known as Mukhanov-Sasaki variable, $\mathcal{Q}_{1}$), density fluctuation on uniform curvature hypersurfaces (given as $\mathcal{M}_{1}$), curvature perturbation on uniform density $\lsp\tilde{\delta\rho_{1}}=0\rsp$ hypersurfaces (given as, $\zeta_{1}$) can be expressed as 
\bea
\nn \mathcal{Q}_{1} &=& \tdvp_1|_{\tilde\psi_1=0} = \delta\vp_{1}+\frac{\vp^{\prime}_{0}}{\mh}\psi_1,\\
\nn \mathcal{M}_{1} &=& \tilde{\delta\rho}_1|_{\tilde\psi_1=0} = \delta\rho_{1}+\frac{\rho^{\prime}_{0}}{\mh}\psi_1,\\
    \zeta_{1}       &=& \tilde\psi_1|_{\tilde{\delta\rho}_1=0} = \psi_1+\frac{\mh}{\rho^{\prime}_{0}}\delta\rho_{1}.
    \label{ug:qmz}
\eea 
It is important to notice that, in the above cases, $\alpha_{1}$ is chosen in such a way that we can get uniform curvature, uniform density density fluctuation and uniform field fluctuation hypersurfaces. In all the cases, $\beta_{1}$ will be given by Eq.(\ref{ugc:f1}). But, the construction of the above gauge invariant quantities is independent of $\beta_{1}$. Therefore, in the observational context, unimodular gravity and GR are identical.
\\
\\
\underline{\emph{Construction of gauge invariant Newtonian potential ($\phi_1$) and curvature ($\psi_1$)}}
\\

As mentioned earlier, due to the additional unimodular constraint, one can not construct gauge invariant definitions $\phi_1$ and $\psi_1$ using longitudinal Newtonian gauge (where $\tilde{B_1}=\tilde{E_1}=0$). Let us assume that we set only one condition, for example $\tilde{E_1}=0$. It thus fixes $\beta_{1}=E_{1}$, uniquely. The unimodular constraint (\ref{ugc:f1}) consequently gives us 
\be
\alpha_{1}^{\prime}+4\mh\alpha_{1}=-\Delta E_{1}.
\label{al:sn}
\ee
The solution of $\alpha_{1}$ can be expressed as
\be
\alpha_{1}=a^{-4}\int a^{4}\lsp-\Delta E_{1}\rsp d\eta + a^{-4}d\lsp\vec{x}\rsp,
\label{al:sn1}
\ee
where $d\lsp\vec{x}\rsp$ is an arbitrary spatial integration constant. Substituting in Eq.(\ref{tt:metric}), one can construct gauge invariant definitions of curvature and Newtonian potential respectively as
\bea
\nn \tilde{\psi_1}&=&\psi_1+\mh\lsb a^{-4}\int a^{4}\lsp-\Delta E_{1}\rsp d\eta + a^{-4}d\lsp\vec{x}\rsp\rsb,\\
\tilde{\phi_1}&=&\phi_1+3\mh\lsb a^{-4}\int a^{4}\lsp-\Delta E_{1}\rsp d\eta + a^{-4}d\lsp\vec{x}\rsp\rsb+\Delta E_{1}.
\label{cvnp:f1}
\eea
Due to the presence of $d\lsp\vec{x}\rsp$, the above definition of $\phi$ and $\psi$ can render spurious gauge modes. Hence, these definitions of gauge invariant curvature and Newtonian potential are not very useful.

%However, the problem of fictitious gauge modes can be avoided, thanks to the linear relations in first order perturbation. One can take linear combinations of the relations given in (\ref{tt:metric}) and construct gauge invariant variables. From the expressions of $\tilde{B1}$ and $\tilde{E1}$ one can eliminate $\beta_{1}$. This gives the expression of $\alpha_{1}$ as 
However, in the case of UG one can calculate the gauge invariant Newtonian potential and curvature on the hypersurfaces where $\tilde{B}_{1}=\tilde{E}_{1}^{\prime}$. On this hypersurface, $\alpha_1$ can be determined as 
\be
\alpha_{1}= - \lsp B_1-E_1^{\prime}\rsp. 
\ee
Substituting this expression back in $\tilde{\psi_1}$ and $\tilde{\phi_1}$, one can construct the gauge invariant curvature and Newtonian potential as
\bea
\nn \label{PSI1PHI1} \Psi_1 &=&\psi_1 - \mh\lsp B_1-E_1^{\prime}\rsp,\\
    \Phi_1 &=&\phi_1 + \mh\lsp B_1-E_1^{\prime}\rsp + \lsp B_1-E_1^{\prime}\rsp^{\prime}.
\eea
Therefore, in first order perturbation, all gauge invariant quantities in GR can also be obtained for UG. The only difference between GR and UG is the choice of hypersurfaces. As in the context of UG one can only set either $B_1$ or $E_1$ equal to zero, then $\Psi_1\neq\psi_1$ and $\Phi_1\neq\phi_1$. Still in this case, like in GR, the relation between the gauge invariant quantities, $\Phi_1=\Psi_1$, holds. The evolution equations of $\Phi_1$ or $\Psi_1$ are also identical in UG and GR \cite{Gao:2014nia}. 

However, it is interesting to note that in UG, if one sets $B_1=E_1^{\prime}\neq 0$, one can obtain $\Phi_1 = \phi_1$ and $\Psi_1 = \psi_1$. Then, from equation (\ref{tf:eep1}) we get $\phi_1=\psi_1$ or $\Phi_1=\Psi_1$, which are identical to Newtonian gauge in GR. Finally, using (\ref{tf:eep2}) and (\ref{tf:eep3}) one obtains the following equation of motion for gravitational potential $\phi_1$
\be
\phi_1^{\prime\prime} + 2\lsp\mh-\frac{\vpb^{\prime\prime}}{\vpb^{\prime}}\rsp\phi_1^{\prime} - \lsb \Delta - 2\lsp \mh^{\prime} - \mh\frac{\vpb^{\prime\prime}}{\vpb^{\prime}}\rsp\rsb\phi_1=0,
\ee
which is also identical to the Newtonian gauge in GR. It is important to note that, to derive the equation of motion for $\phi_1$, the unimodular constraint equation is not used.

%$$$$$$$$$$$$$$$$$$$$$$$$$$$$$$$$$$$$$$$$$$$$$$$$$$$$$$$$$$$$$$$$$$$$$$$$$$$$$$$$$$$$
\subsection{Evolution of first order Mukhanov-Sasaki variable, $\mathcal{Q}_1$}

On the uniform curvature hypersurfaces $\lsp\tilde{\psi}_1=0\rsp$, the expression of $\alpha_1$ is given by 
\be
\alpha_1 = -\frac{\psi_1}{\mh}.
\label{al1:uch}
\ee
\noindent and the Klein-Gordon equation takes the following form,
\be
\tdvp^{\prime\prime}_1 + 2\mh\tdvp^{\prime}_1 - \Delta\tdvp_1 + a^{2}V_{, \vp \vp}\tdvp_1 - \vp^{\prime}_0\tilde{\phi_1}^{\prime} + 2 a^{2} V_{, \vp}\tilde{\phi_1} - \vpb^{\prime} \Delta\lsp\tilde{B_1} - \tilde{E_1}^{\prime}\rsp = 0.
\label{kg1:uch}
\ee
Here, the field fluctuation on the uniform curvature hypersurfaces is defined as $\tdvp_1 = \dvp_1 + \frac{\vpb^{\prime}}{\mh}\psi_1 = \mathcal{Q}_1$. The unimodular constraint on this hypersurface becomes,
\be
\Delta\tilde{E_1} + \tilde{\phi_1} = 0.
\label{ugc1:uch}
\ee
From the $\lsp i\neq j\rsp$ components of the perturbed field equations (\ref{tf:etp}), using the background equations (\ref{back:frd}), we get  
\be
\tilde{B_1}^{\prime} - \tilde{E_1}^{\prime \prime} = -2\mh \lsp\tilde{B_1} - \tilde{E_1}^{\prime}\rsp - \tilde{\phi_1}.
\label{feij:uch}
\ee
Using (\ref{feij:uch}), one can obtain the $\lsp\eta, \eta\rsp$ component of the perturbed field equations on the same hypersurfaces
\be
\mh\tilde{\phi_1}^{\prime} - \mh\Delta\lsp\tilde{B_1} - \tilde{E_1}^{\prime}\rsp = 8\pi G \vpb^{\prime}\tdvp^{\prime}_1.
\label{feee:uch}
\ee
Finally, the $\lsp\eta, i\rsp$ component can be written as
\be
\mh\tilde{\phi_1} = 4\pi G  \vpb^{\prime}\tdvp_1.
\label{feei:uch}
\ee

From (\ref{feei:uch}) one can obtain the expression of the Newtonian potential on the uniform curvature hypersurfaces in terms of the field fluctuation $\lsp\tdvp_1\rsp$
\be
\tilde{\phi_1} = 4\pi G  \frac{\vpb^{\prime}}{\mh}\tdvp_1.
\label{feei:uch1}
\ee
Using Eqs.~(\ref{ugc1:uch}), (\ref{feee:uch}) and (\ref{feei:uch1}), $\tb_1$ and $\te_1$ are
\be
\tilde{B_1} = -8\pi G \frac{\vpb^{\prime}}{\mh} \Delta^{-1}\lsp\tdvp^{\prime}_1\rsp, \qquad
\tilde{E_1} = -4\pi G \frac{\vpb^{\prime}}{\mh} \Delta^{-1}\lsp\tdvp_1\rsp.
\label{tbte1:uch}
\ee
Finally, using the above expressions in (\ref{kg1:uch}), one can write the evolution equation for the Mukhanov-Sasaki variable $\lsp\mathcal{Q}_1\rsp$ as 
\be
\mathcal{Q}_1^{\prime\prime} + 2\mh\mathcal{Q}_1^{\prime} - \Delta\mathcal{Q}_1 + \lsb a^{2} V_{, \vp \vp} + 8\pi G \lsc a^{2} V_{,\vp}\frac{\vpb^{\prime}}{\mh} -\vpb^{\prime}\lsp\frac{\vpb^{\prime}}{\mh}\rsp^{\prime}\rsc \rsb\mathcal{Q}_1=0.
\label{kg1:uch2}
\ee
Using the background equations, it can be checked that the evolution equation of $\mathcal{Q}_1$ in UG is identical to GR (see for instance Refs.~\cite{Malik:2005cy, Malik:2006ir}). In the first order perturbation, one can note that the expression of $\tp_1$ is identical in GR and UG. On the other hand, $\tb_1$ and $\te_1$ are individually different in UG. But, in UG, the linear combination $\lsp\tb_1 - \te_1^{\prime}\rsp$ remains same as in GR. It is interesting to note that, in perturbed Klein Gordon equation (\ref{kg1:uch}), the terms involving $\tb_1$ and $\te_1$ can be written in terms of the linear combination $\lsp\tb_1 - \te_1^{\prime}\rsp$. Therefore, the governing equation of $\mathcal{Q}_1$ is same in both GR and UG. 

It should be noticed that the derivation of the equation of first order Mukhanov-Sasaki variable does not require the unimodular constraint equation (\ref{ugc1:uch}). The first order unimodular constraint is only used to find the expression of $\te_1$ in terms of $\tdvp_1$ or $\mathcal{Q}_1$.

It should also be noticed that in case of GR, setting $\te_1=0$ or $\beta_1=E_1$ is a choice. In GR, one can make a different choice for $\te_1$ other than zero. As long as Eq.~(\ref{feee:uch}) is satisfied, equation of motion for $\mathcal{Q}_1$ remains unchanged even in GR. Thus, in GR, the equation of motion of $\mathcal{Q}_1$ (\ref{kg1:uch2}) remains the same for any choice of $\te_1$. In UG, only one particular choice of $\te_1$ is allowed, which is given by (\ref{ugc1:uch}). However this particular choice of $\te_1$ does not have any effect on the equation of motion of $\mathcal{Q}_1$ in UG. In the next Section, we will show the equivalence between GR and UG based on first order curvature perturbation on uniform density hypersurfaces. 

\subsection{Curvature perturbation on uniform density hypersurfaces, $\zeta_1$}

The gauge invariant definition of the curvature perturbation in uniform density hypersurfaces is given as 
\be
\zeta_1=\psi_1+\mh\frac{\delta\rho_1}{\rho_0^{\prime}}=\psi_1-\frac{\delta\rho_1}{3\lsp\rho_0 +p_0\rsp},
\label{cp:udh}
\ee
where $\rho_0$ and $p_0$ are the background energy density and pressure density. In the last expression, we have used the conservation of energy density given by equation (\ref{cont1}). In case of GR, without cosmological constant, it is known that 
\be
\delta\rho_{1}|_{GR}=\frac{1}{4\pi Ga^{2}}\lsb \Delta\psi_1-3\mh\lsp\psi_1^{\prime}+\mh\phi_1\rsp\rsb. 
\label{dr:gr}
\ee
In the above equation, we have used the following expression
\be
\delta G^{\eta}_{\eta(1)}|_{GR}=\frac{2}{a^{2}}\lsb \Delta\psi_1-3\mh\lsp\psi_1^{\prime}+\mh\phi_1\rsp\rsb.
\label{gee:gr}
\ee
Therefore, in GR, the expression of $\zeta_1|_{GR}$ in (\ref{cp:udh}) becomes
\be
\zeta_1|_{GR}\approx\psi_1+\frac{\mh}{4\pi Ga^2}\frac{\psi_1^{\prime}+\mh\phi_1}{\lsp \rho_0+p_0\rsp},
\label{z:gr} 
\ee
on large scale ($\Delta\rightarrow 0$). Using (\ref{tf:eep2}), which is same in GR, and the background equations, one gets $\zeta_1|_{GR}\approx \mr_1|_{GR}$ in large scales.

In contrast, in UG, the perturbed field equation can be written as
\be
\delta G^{\mu}_{\nu(1)}|_{GR}+\delta^{\mu}_{\nu}\delta\hat{\Lambda}_1=8\pi G\delta T^{\mu}_{\nu(1)}|_{UG},
\label{pfe:ugr}
\ee
where $\delta\hat{\Lambda}_1=\frac{1}{4}\lsp\delta R_1+8\pi G\delta T_1\rsp$.
Therefore, in unimodular gravity, the expression of perturbed energy density becomes
\be
\delta\rho_1|_{UG}=\frac{1}{8\pi G}\lsb\delta G^{\eta}_{\eta(1)}|_{GR}+\delta\hat{\Lambda}_1\rsb.
\label{dr:ugr}
\ee
Using equation (\ref{dr:ugr}) and the expression of $\delta\rho_1|_{UG}$, the expression of curvature perturbation in uniform-density hypersurfaces in UG can be obtained from (\ref{cp:udh}):
\be
\zeta_1|_{UG}=\psi_1-\frac{1}{8\pi G}\frac{\delta G^{\eta}_{\eta(1)}|_{GR}+\delta\hat{\Lambda}_1}{3\lsp \rho_0+p_0\rsp}.
\label{z:ugr1}
\ee
Finally, using (\ref{gee:gr}), the expression of curvature perturbation in uniform-density hypersurfaces on large scales is approximated by 
\be
\zeta_1|_{UG}\approx\psi_1+\frac{\mh}{4\pi Ga^2}\frac{\psi_1^{\prime}+\mh\phi_1}{\lsp \rho_0+p_0\rsp}-\frac{1}{8\pi G}\frac{\delta\hat{\Lambda}_1}{3\lsp \rho_0+p_0\rsp}.
\label{z:ugr3}
\ee

\noindent\underline{\emph{The evolution of curvature perturbation in large scale}} ---
\\
The time derivative of Eq. (\ref{cp:udh}) gives us
\be
\zeta_1^{\prime}=\psi_1^{\prime}-\lsb\frac{\delta\rho_1}{3\lsp\rho_0+p_0\rsp}\rsb^{\prime},
\label{zp:ugr}
\ee
where the background continuity equation (\ref{cont1}) is used. First order perturbation of the energy conservation equation on the large scale gives us \cite{Wands:2000dp,Lyth:2003im}
\be
\delta\rho_1^{\prime}\approx-3\mh\lsp\delta\rho_1+\delta p_1\rsp+3\lsp\rho_0+p_0\rsp\psi_1^{\prime},
\label{contp1}
\ee
which eventually gives us the large scale time evolution of curvature perturbation as
\be
\zeta_1^{\prime}\approx\frac{\mh}{\lsp\rho_0+p_0\rsp}\lsp\delta p_1-\frac{p_0^{\prime}}{\rho_0^{\prime}}\delta\rho_1\rsp.
\label{zp:ugr1}
\ee
The above expression of large-scale time-evolution of the curvature perturbation (\ref{zp:ugr1}) is valid in both UG and GR.

Using the continuity equation (\ref{cont1}) and the Klein-Gordon equation of the scalar field (\ref{kg}), we get 
\be
\frac{p_0^{\prime}}{\rho_0^{\prime}}=-\frac{\vp_0^{\prime\prime}}{\mh\vp_0^{\prime}}=\delta-1,
\ee
where $\delta$ is the slow-roll parameter. This gives us 
\be
\zeta_1^{\prime}\approx\frac{\mh}{\lsp\rho_0+p_0\rsp}\lsb\lsp\delta\rho_1+\delta p_1\rsp-\lsp\delta\rsp\delta\rho_1\rsb.
\label{zp:ugr2}
\ee
From equation (\ref{pfe:ugr}), the following relations can be obtained for unimodular gravity
\bea
\lsp\delta\rho_1+\delta p_1\rsp|_{UG} &=& \frac{1}{8\pi G}\lsp\delta G^{\eta}_{\eta(1)}|_{GR}-\delta G^{i}_{i(1)}|_{GR}\rsp=\lsp\delta\rho_1+\delta p_1\rsp|_{GR},\label{zp:ugr3}\\
\lsp\delta\rsp\delta\rho_1|_{UG} &=& \frac{\lsp\delta\rsp}{8\pi G}\lsp\delta G^{\eta}_{\eta(1)}|_{GR}+\delta\hat{\Lambda}_1\rsp=\lsp\delta\rsp\delta\rho_1|_{GR}+\lsp\delta\rsp\frac{\delta\hat{\Lambda}_1}{8\pi G}, \label{zp:ugr4}
\eea
where $\delta\rho_1=\delta T^{\eta}_{\eta(1)}$ and $\delta p_1=-\delta T^{i}_{j(1)}\delta_{ij}$. Finally,
the relation between the evolution of $\zeta_1$ in GR and UG can be expressed as
\be
\zeta^{\prime}_1|_{UG}=\zeta^{\prime}_1|_{GR}-\frac{\mh}{\lsp\rho_0+p_0\rsp}\lsp\delta\rsp\frac{\delta\hat {\Lambda}_1}{8\pi G}.
\ee

On the super-Hubble scale, one can show that $\delta R_1\approx\frac{48\pi G}{a^{2}}\lsp\mh\vp_0^{\prime}+\vp_0^{\prime\prime}\rsp\delta\vp_1$ and $8\pi G\delta T_1\approx-\frac{48\pi G}{a^{2}}\lsp\mh\vp_0^{\prime}+\vp_0^{\prime\prime}\rsp\delta\vp_1$. Therefore, $\delta\hat{\Lambda}_1=\delta R_1+8\pi G\delta T_1=0$. It can thus be concluded that $\zeta_1|_{UG}=\zeta_1|_{GR}$ and $\zeta^{\prime}_1|_{UG}=\zeta^{\prime}_1|_{GR}$. 

It can also be shown that the evolution of $\mathcal{M}_1$ is same in GR and UG. As a consequence, GR and UG are identical in the first order perturbation. 
%%%%%%%%%%%%%%%%%%%%%%%%%%%%%%%%%%%%%%%%%%%%%%%%%%%%%%%%%%%%%%%%%%%%%%%%%%%%%%%%%%%%%%%%%%%%%%%%%%%%%

\section{Sachs-Wolfe effect and CMB temperature fluctuation in UG}\label{SW_CMB}

In this Section, we derive collision-less Boltzmann equation of CMB photons at first order for super-Hubble modes to obtain the main contribution of the CMB temperature anisotropies at large scale \textit{i.e.}, the Sachs-Wolfe effect. We follow Refs.~\cite{Gao:2014nia, Mukhanov:1992ab,Mukhanov:2004cmb}. For the gauge choice $E_1=0$ and $B_1\neq0$, equations (\ref{PSI1PHI1}) reduce to
\begin{eqnarray}
\nn\Psi_1 &=& \psi_1 - \mathcal{H} B_1,\\
\Phi_1 &=& \phi_1 + \mathcal{H} B_1+B^{'}_1.
\end{eqnarray}

During recombination, as more and more hydrogen atoms form and as the
temperature decreases, photons no longer interact with matter. Afterwards, the CMB
photons can be described as a gas of non-interacting identical particles described by kinetic equations. 
By solving the Boltzmann equation for freely propagating radiation, the relationship between the CMB temperature fluctuations and the gravitational potential can be found. 

\subsection{Colisionless Boltzmann equation}

First of all, let us consider the collision-less Boltzmann equation for the photon distribution function $f$
\begin{eqnarray}\label{Boltzmann}
\frac{\partial f}{\partial \eta}+\frac{d x^i}{d \eta}\frac{\partial f}{\partial x^i}+\frac{d p_j}{d \eta}\frac{\partial f}{\partial p_j}=0,
\end{eqnarray}

\noindent where $\frac{dx_i}{d \eta}$ and $\frac{dp_i}{d \eta}$ are the derivatives calculated along the geodesics. For an observer with 4-velocity $u^\alpha$ in an arbitrary coordinate system and a photon with 4-momentum $p_\alpha$, the frequency of this photon measured by the observer is given by
\begin{eqnarray}
\omega = u^\alpha p_\alpha.
\end{eqnarray}

For massless photons, $p_\alpha p^\alpha = 0$. 
The distribution function of the non-interacting photons is given by the function $f$
\begin{eqnarray}\label{pdf}
f=\frac{2}{{\rm exp}[\omega/T(\eta,x^i,l^i)]-1},
\end{eqnarray}

\noindent which depends on the direction $l^i$ and on the location of the observer $x^i$ at the moment
of conformal time $\eta$. In a nearly isotropic universe, the temperature  can be decomposed as
\begin{eqnarray}
T(x^\alpha,l^i)=T_0 (\eta)+\delta T(x^\alpha,l^i),
\end{eqnarray}

\noindent where $T_0$ is the average background temperature and $\delta T$ temperature fluctuation. As a consequence, up to the first order in the metric perturbations, one obtains
\begin{eqnarray}\label{momentum}
p^\eta = \frac{p}{a^2} \left(1 + \psi_1 - \phi_1 \right), \qquad
p_\eta = p (1 + \psi_1 + \phi_1 - l^i B_{1 ,i} ) ,
\end{eqnarray}

\noindent where we have introduced $p= |p| = \sqrt{\left(Σp_i^2\right)}$. Then, the geodesics equations are given by
\begin{eqnarray}
\frac{d x^\alpha}{d \lambda}=p^\alpha,\qquad
\frac{d p_\alpha}{d \lambda}=\frac{1}{2}\frac{\partial g_{\gamma \delta}}{\partial x^\alpha} p^\gamma p^\delta,\label{geodesics2}
\end{eqnarray}

\noindent where $\lambda$ is an affine parameter along the geodesic. 

For the background, using Eqs. (\ref{momentum}), (\ref{geodesics2}) and (\ref{pdf}), the Boltzmann equation (\ref{Boltzmann}) yields
\begin{eqnarray}
(aT)^{'}=0.
\end{eqnarray}
Thus, we obtain the well-known relation that in a homogeneous universe the temperature is inversely proportional to the scale factor. Then, expanding Boltzmann equation to first order in cosmological perturbations, we get
\begin{eqnarray}
\left (\frac{\partial }{\partial \eta} +l^i \frac{\partial }{\partial x^i}\right ) \left ( \phi_1+\frac{\delta T}{T_0}+B^{'}_1\right ) =\frac{\partial }{\partial \eta}\left(2\Phi_1\right).
\label{sw:ug}
\end{eqnarray}

Like in GR, as $\Phi_1$ is constant on large scales \citep{Gao:2014nia}, the final expression of perturbed Boltzmann equation (\ref{sw:ug}) gives us
\begin{eqnarray}
\phi_1+\frac{\delta T}{T_0}+B^{'}_1=\mathrm{constant},
\label{sw:ug2}
\end{eqnarray}
along null geodesics. As a consequence, if we evaluate this constant today and at the time of last scattering, we deduce the following equality
\begin{eqnarray}
\frac{\delta T}{T_0}|_{\rm now} &=&\frac{\delta T}{T_0}|_{\rm LSS}+\phi_1|_{\rm LSS}+B^{'}_1|_{\rm LSS}+{\rm monopole~contributions}.
\end{eqnarray} 

In Ref. \cite{Gao:2014nia}, the authors claimed that the equation (\ref{sw:ug2}) is different compared to GR, as $B_1 \neq 0$ in UG. But, we argue that, in GR setting $B_1=0$ is a choice which gives us $\phi_1+\frac{\delta T}{T_0}=\mathrm{constant}$. If, in GR we set $B_1 \neq 0$, perturbed Boltzmann equation yields the same expression as (\ref{sw:ug2}). Therefore, it can be concluded that first order perturbed Boltzmann equations are same in GR and UG. In the following we show that, even though $B_1\neq 0$ in UG,  the expression of $\frac{\delta T}{T_0}$ is identical with GR.

\subsection{Kinetic energy-momentum tensor}

The kinetic energy-momentum tensor, which characterizes the gas of free photons after decoupling is expressed as
\begin{eqnarray}
T_{\beta}^{\alpha} = \frac{1}{\sqrt{-g}}\int d^3 p \frac{p^\alpha p^\beta}{p^0} f.
\end{eqnarray}
Substituting the perturbed metric into the above equation and taking a Planckian distribution, the ($\eta,\eta$) and the ($\eta,i$) components of the kinetic energy-momentum tensor are obtained as following
\begin{eqnarray}
T^{\eta}_\eta &=&\rho_\gamma\left(1+\delta_\gamma\right) \approx {\left(T_0\right)}^{4}\int dy d^2 l \left(1+4\frac{\delta T}{T_0}+3l^i B_{1,i}\right)y^3 f(y),\\   
T^{i}_\eta &\approx&  {\left(T_0\right)}^{4} \int dy d^2 l\lsb \lsp 1 + 4\psi_1\rsp l^{i} + \left(4l^i \frac{\delta T}{T_0}+3l^i l^j B_{1,j}-B_{1,i}\right)\rsb y^3 f(y),
\end{eqnarray}
where $y=\frac{\omega}{T}$. The fractional radiation density perturbation is defined as $\delta_\gamma=\frac{\delta \rho_\gamma}{\rho_\gamma}$, where $\rho_\gamma=4\pi{\left(T_0\right)}^{4}\int dy~f(y)y^3$ is the energy density of photons after recombination and $\delta\rho_\gamma$ is the energy density fluctuation of photon. We deduce from the latter equations
\begin{eqnarray}\label{delta-gamma}
 \delta_\gamma     &=& \int \frac{d^2 l}{4\pi} \left(4\frac{\delta T}{T_0}+3l^i B_{1,i}\right),\\\label{delta-gamma-prime}
 \delta_\gamma^{'} &=& \int \frac{d^2 l}{4\pi} \left(-4l^i \partial_i\frac{\delta T}{T_0}-3l^i \partial_iB_{1,j}l^j +\partial_i B_{1,i}\right).
%\delta_\gamma^{'} &=& \int \frac{d^2 l}{4\pi} \left(-4l^i \nabla_i\frac{\delta T}{T_0}-3l^i %\nabla_iB_{1,j}l^j +\nabla_i B_{1,i}\right).
\end{eqnarray}
To obtain the expression of $\delta_{\gamma}^{'}$, we use the perturbed fluid equation, $T^{i}_{\eta ,i} =\lsp 4\psi_1 - \delta_{\gamma} \rsp^{'} \rho_{\gamma}$. For further details, one can check Refs. \cite{Gao:2014nia, Mukhanov:1992ab, Mukhanov:2004cmb}.

\subsection{Sachs-Wolfe effect}

The solution that satisfy both equations Eqs. (\ref{delta-gamma}) and (\ref{delta-gamma-prime}) in Fourier space is 
\begin{eqnarray}\label{deltaT-LSS-1}
 \left(\frac{\delta T}{T_0}\right)_k |_{\rm LSS} = \frac{1}{4}\lsp\delta_{\gamma}\rsp_k + k_j l^j\frac{3\iota}{4k^2} \lsp\delta^{'}_{\gamma}\rsp_k,
\end{eqnarray} 
where $\iota$ is the imaginary square root of ($-1$).

It can be checked, by taking the Fourier transform of Eqs. (\ref{delta-gamma}) and (\ref{delta-gamma-prime}) and replacing in the two integrals $\left(\frac{\delta T}{T_0}\right)_k |_{\rm LSS}$ by the expression given in (\ref{deltaT-LSS-1}), that this solution is consistent.

Eq. (\ref{deltaT-LSS-1}) has already been derived in \cite{Mukhanov:1992ab,Mukhanov:2004cmb} in the GR setting. The solution found in \cite{Gao:2014nia} has an extra term: $-\frac{3}{4}k_j l^j\left(B_1\right)_k |_{\rm LSS}$. Substituting this term in (\ref{delta-gamma-prime}), the right hand side becomes $\lsp\delta^{'}_{\gamma}\rsp_k+\iota k^2 (B_1)_k$ in Fourier modes with $B_1 \neq 0$. This is not consistent with the Fourier transform of the left hand side of (\ref{delta-gamma-prime}), which is only $\lsp\delta^{'}_{\gamma}\rsp_k$. As a consequence, we can say that the solution found in \cite{Gao:2014nia} should not be taken into account. We thus state that, in comparison to GR, there is no modification of the expression of $\lsp\frac{\delta T}{T_0}\rsp_k|_{\rm LSS}$ in the UG case.

The remaining part of the computation can be found in \cite{ Mukhanov:1992ab,Mukhanov:2004cmb}. The final expression in real space describing the temperature fluctuations of the CMB on the last scattering surface is given by
\begin{eqnarray}
  \left(\frac{\delta T}{T_0}\right) |_{\rm now} = \frac{1}{3}\Phi_1 |_{\rm LSS},
  \label{sw:ug3}
\end{eqnarray} 
where we can recognize the well-known Sachs-Wolfe effect which describe a plateau at large scales in the power spectrum of the CMB. We finally underline that we detect no difference between the UG and the GR Sachs-Wolfe contribution unlike \cite{Gao:2014nia} who claims that there is a dipole contribution on large scales. This result is supported by Ref. \cite{peter2009primordial}, where CMB temperature fluctuation is calculate in generic gauge. 

It is possible to derive the above result for the general case $B_1\neq0$ and $E_1\neq 0$. 
Even in the most general case, the UG constraint is only equivalent to a choice of gauge. In the most general case the Sachs-Wolfe effect is also given by (\ref{sw:ug3}). As a consequence there is no way to detect in CMB data if UG or GR is the most favoured theory.

%%%%%%%%%%%%%%%%%%%%%%%%%%%%%%%%%%%%%%%%%%%%%%%%%%%%%%%%%%%%%%%%%%%%%%%%%%%%%%%%%%%%%%%%%%%%%%%%%%%%%

\section{Second order perturbations}\label{second_order}

The aim of this Section is to obtain the second order gauge invariant variable and its equation of motion. Similar to first order perturbation, one can obtain various gauge invariant quantities in second order. In this work we construct second order Mukhanov-Sasaki variable in UG. Then we present the comparison between GR and UG based on the second order Mukhanov-Sasaki variable. To achieve this, we first give the perturbed second order field equations. Finally, using the field equations, we find the equation of motion for second order Mukhanov-Sasaki equation for UG. 

In second order the trace-free part of the Einstein tensor is given as
\be
\delta\hat{G}^{\mu}_{\nu \lsp 2\rsp}=8\pi G \delta\hat{T}^{\mu}_{\nu \lsp 2\rsp},
\label{tf2:eep}
\ee
where $\delta\hat{G}^{\mu}_{\nu \lsp 2\rsp}$ and $\delta\hat{T}^{\mu}_{\nu \lsp 2\rsp}$ are the trace-free part of the perturbed Einstein tensor and energy-momentum tensor respectively.
\subsection{Perturbed trace-free energy-momentum and Einstein tensor}
\underline{\emph{Perturbed trace-free energy-momentum tensor ---}}

The second order ($\eta, \eta$), ($\eta, i$) and ($i, j$) components of the trace-free Einstein equation are given as
\bea
\nn \delta\hat{T}^{\eta}_{\eta (2)} &=& \frac{1}{2a^2}(\frac{3}{2}\, \vp_0^{\prime}\dvp_2^{\prime} + \frac{3}{2}\dvp_1^{\prime 2} - 6 \vp_{0}^{\prime}  \dvp_1^{\prime}\phi_1 - \frac{3}{2}\phi_2 \vp_0^{\prime 2} +
 6 \vp_0^{\prime 2} \phi_1^2 - \frac{3}{2} B_{1 ,i}B_{1 ,i}\vp_0^{\prime 2} + \\
 \nn && \frac{1}{2} \dvp_{1 ,i}\dvp_{1 ,i} - \vp_0^{\prime}B_{1 ,i}\dvp_{1 ,i}),
\eea
\bea
\nn\delta\hat{T}^{\eta}_{i (2)} &=& \frac{1}{2a^2}(\vp_0^{\prime}\dvp_{2 ,i} + 2\dvp_{1}^{\prime}\dvp_{1 ,i} - 4\vp_{0}^{\prime}\dvp_{1 ,i}\phi_1 ),
\eea
\bea
\nn\delta\hat{T}^{i}_{j (2)} &=& \frac{1}{2a^2}[( - \frac{1}{2}\vp_{0}^{\prime}\dvp_{2}^{\prime} - \frac{1}{2}\dvp_{1}^{\prime 2} + 2\varphi_{0}^{\prime}\dvp_{1}^{\prime}\phi_1 + \frac{1}{2} \phi_2 \vp_{0}^{\prime 2} - 2 \vp_{0}^{\prime 2}\phi_1^{2} + \frac{1}{2}B_{1 ,i}B_{1 ,i}\vp_{0}^{\prime 2} + \\
 && \vp_{0}^{\prime}B_{1 ,i}\dvp_{1 ,i} + 
\frac{1}{2}\vp_{1 ,i}\vp_{1 ,i} ) {\delta}_{i j} - 2 \vp_{0}^{\prime}B_{1 ,i}\dvp_{1 ,j} - 2 \dvp_{1 ,i}\dvp_{1 ,j}].
\label{em2:ug}
\eea
Here, $\dvp_2$ is the second order fluctuation of the of the scalar field $\vp$.
\\~\\
\noindent\underline{\emph{Perturbed trace-free Einstein tensor ---}}

The second order ($\eta, \eta$), ($\eta, i$) and off-diagonal ($i,j$) components of trace-free Einstein tensor are respectively 
\bea
\nn \delta\hat{G}^{\eta}_{\eta (2)} &=&\frac{1}{a^2} ( - 3\mh^2 B_{1 ,i}B_{1 ,i} + \frac{3}{2}\frac{a^{\prime\prime}}{a}B_{1 ,i}B_{1 ,i} - \frac{1}{4}\Delta\lsp\mh B2\rsp + \mh B_{1 ,ij}E_{1 ,ij} + \frac{1}{2}\mh B_{1 ,i}\Delta E_{1 ,i} + \\
\nn && \frac{1}{4}E_{1 ,ij}^{\prime}E_{1 ,ij}^{\prime} - 6\mh\phi_1 \phi_1^{\prime} - \mh E_{1 ,ij}E_{1 ,ij}^{\prime} + \frac{1}{4}\mh \Delta E_2^{\prime} + \frac{1}{2} \mh \phi_{1 ,i} B_{1 ,i} + \phi_1 \Delta \lsp\mh B_1\rsp - \\
\nn && \phi_1\Delta B_1^{\prime} - \phi_1\Delta\phi_1 + \phi_1\Delta E_1^{\prime\prime} - \phi_1\Delta\lsp\mh E_1^{\prime}\rsp - 3\mh^2 \phi2 + 12\mh^2 \phi_1^2 + \frac{1}{4}\Delta B_1\Delta B_1 - \\
\nn && \frac{1}{4}\Delta E_{1 ,j}\Delta E_{1 ,j} - \frac{1}{2}\Delta B_1\Delta E_1^{\prime} - \frac{1}{4}B_{1, ij}B_{1 ,ij} + \frac{1}{4}E_{1 ,ijk}E_{1 ,ijk} + \frac{1}{4}\Delta E_1^{\prime}\Delta E_1^{\prime} +\\
\nn && \frac{1}{2}B_{1 ,ij}E_{1 ,ij}^{\prime} + \frac{3}{2}\mh B_{1 ,i}B_{1 ,i}^{\prime} - \frac{1}{2}\phi_1^{\prime}\Delta B_1 - E_{1 ,ij}B_{1 ,ij}^{\prime} - \frac{1}{2}B_{1 ,i}^{\prime}\Delta E_{1 ,i} + \frac{1}{4}\Delta B_2^{\prime} - \\
\nn && \phi_{1 ,ij}E_{1 ,ij} - \frac{1}{2}\phi_{1 ,i}\Delta E_{1 ,i} + \frac{1}{4}\Delta\phi_2 + E_{1 ,ij}E_{1 ,ij}^{\prime\prime} - \frac{1}{4}\Delta E_2^{\prime\prime} + \frac{3}{4}\mh\phi_2^{\prime} + \frac{1}{2} \phi_1^{\prime}\Delta E_1^{\prime} - \\
\nn && \frac{1}{2} \phi_{1 ,i} \phi_{1 ,i} + \frac{3}{2} \frac{a^{\prime\prime}}{a} \phi_2 - 6\frac{a^{\prime\prime}}{a} \phi_1^{2}),
\eea
\bea
\nn \delta\hat{G}^{\eta}_{i (2)} &=& \frac{1}{a^2} ( - E_{1 ,ij}^{\prime}\Delta E_{1 ,j} - \phi_{1 ,i}\Delta B_1 + E_{1 ,jk}^{\prime}E_{1 ,jki} - 8\mh\phi_1\phi_{1 ,i} + \mh\phi_{2 ,i} - \phi_{1 ,j}E_{1 ,ij}^{\prime} +\\
\nn && \phi_{1 ,i}\Delta E_1^{\prime} + \phi_{1 ,j}B_{1 ,ji} + 2\mh B_{1 ,j} B_{1 ,ji}),
\eea
\bea
\nn \delta\hat{G}^{i}_{j (2)} &=& \frac{1}{a^2}(- 2\mh B_{1 ,i} \phi_{1 ,j} - 2\phi_1 B_{1 ,ij}^{\prime} - \phi_1^{\prime} B_{1 ,ij} + \frac{1}{2} B_{2 ,ij}^{\prime} -  B_{1 ,k}^{\prime} E_{1 ,ijk} - \Delta B_1  B_{1 ,ij} + \\
\nn && E_{1 ,ijk}\Delta E_{1 ,k} + 4\mh\phi_1 E_{1 ,ij}^{\prime} + 2\phi_1 E_{1 ,ij}^{\prime\prime} +  \phi_1^{\prime} E_{1 ,ij}^{\prime} - \mh E_{2 ,ij}^{\prime} - \frac{1}{2} E_{2 ,ij}^{\prime\prime} - \\
\nn && 2\mh B_{1 ,k} E_{1 ,ijk} + \Delta B_1 E_{1 ,ij}^{\prime} - 2\phi_1 \phi_{1 ,ij} -  \phi_{1 ,i} \phi_{1 ,j} + \frac{1}{2} \phi_{2 ,ij} +  B_{1 ,ik} B_{1 ,jk} - \\
\nn &&  E_{1 ,ikl} E_{1 ,jkl}
 - 4\mh\phi_1 B_{1 ,ij} + \mh B_{2 ,ij} + B_{1 ,ij}\Delta E_1^{\prime} -  \phi_{1 ,k} E_{1 ,ijk} - E_{1 ,ij}^{\prime}\Delta E_1^{\prime} - \\
 \nn && B_{1 ,jk} E_{1 ,ik}^{\prime} - B_{1 ,ik} E_{1 ,jk}^{\prime} + 2 E_{1 ,ik}^{\prime} E_{1 ,jk}^{\prime} - 2 E_{1 ,ik} B_{1 ,jk}^{\prime} + 4\mh E_{1 ,ik} E_{1 ,jk}^{\prime} + \\
  && 2 E_{1 ,ik} E_{1 ,jk}^{\prime\prime} - 2 E_{1 ,ik} \phi_{1 ,jk} - 4\mh B_{1 ,jk} E_{1 ,ik}), \qquad \lsp i\neq j\rsp.
  \label{ee2:ug}
\eea
In this Section, we set $\psi_1=\psi_2=0$. This will be useful when we use uniform curvature hypersurfaces. In the uniform curvature hypersurfaces, all the above equations can be written by replacing the variables with tilde on top of them.

\subsection{Analysis of the second order perturbation}

On the uniform curvature hypersurfaces, ($\eta, i$) components of the Einstein equation can be written as %(here we put `tilde' on the variables to denote uniform curvature hypersurfaces)
\bea
\nn && \mh\tilde\phi_{2 ,i} - 4\mh\tilde\phi_1\tilde\phi_{1 ,i} + 2\mh\tilde B_{1 ,j}\tilde B_{1 ,ij} + \tilde\phi_{1 ,j}\tilde B_{1 ,ij} - \tilde\phi_{1 ,i}\Delta\tilde B_1 -\tilde E_{1 ,ij}^{\prime}\Delta\tilde E_{1 ,j} + \tilde E_{1 ,jk}^{\prime}\tilde E_{1 ,jki} -\\
&& \tilde\phi_{1, j}\tilde E_{1 ,ij}^{\prime} + \tilde\phi_{1 ,i}\Delta\tilde E_{1}^{\prime} = 4\pi G \lsp \vpb^{\prime}\tdvp_{2 ,i} + 2\tdvp_1^{\prime}\tdvp_{1 ,i}\rsp.
\label{eeei:ug1}
\eea
Taking the divergence of (\ref{eeei:ug1}) and then computing the trace, the above equation can be written as 
\be
\mh\tilde\phi_2 = 4\pi G \vpb^{\prime}\tdvp_2 + \lsb 8\pi G\tdvp_1\tdvp_1^{\prime} + 2\mh\tilde\phi_1^2 - \mh\lsp\tilde B_{1 ,j}\tilde B_{1 ,j}\rsp\rsb + \Delta^{-1} X_2, 
\label{eeei:ug2}
\ee
where $\Delta^{-1}$ is the inverse of the Laplacian operator, and $X_2$ is a function of first order quantities defined as
\bea
\nn X_2 &=& \Delta\tilde\phi_1\Delta\lsp\tilde B_1-\tilde E_1^{\prime}\rsp - \tilde\phi_{1 ,ij}\lsp\tilde B_1-\tilde E_1^{\prime}\rsp_{,ij} - \Delta\tilde E_{1 ,i}^{\prime}\Delta\tilde E_{1 ,i} -\tilde E_{1 ,ijk}^{\prime}\tilde E_{1 ,ijk} -\\
&& 8\pi G \lsp \tdvp_1\tdvp_{1 ,i}^{\prime}\rsp_{,i}.
\label{X2}
\eea
In the above expression, the right hand side can be expressed entirely in terms of $\tdvp_1$ and its derivatives. This is identical to the expression obtained in GR. The only difference is that in GR one can set $\tilde E_{1}=0$. However, in case of UG, $\tilde E_1$ is given by equation (\ref{ugc1:uch}). 

On the uniform curvature hypersurfaces, the off-diagonal ($i,j$) components of the field equation can be written as
\bea
\nn (\tilde B_2^{\prime} -\tilde E_2^{\prime\prime}) + 2\mh ( \tilde B_2-\tilde E_2^{\prime}) + \tilde \phi_2 &=& \partial_{i}^{-1}\partial_{j}^{-1} [ 2(\tilde B_1-\tilde E_1^{\prime})_{,i j} \{\tilde \phi_1^{\prime} + \Delta(\tilde B_1-\tilde E_1^{\prime})\} + 2\tilde \phi_{1 ,i}\tilde \phi_{1 ,j} - \\
\nn && 2 ( \tilde B_1-\tilde E_1^{\prime})_{, ik}( \tilde B_1-\tilde E_1^{\prime})_{, jk} - 2 \tilde E_{1 ,ik}^{\prime}\tilde E_{1 ,jk}^{\prime} + \\ \nn && 2\tilde E_{1 ,ijk}(\tilde B_1^{\prime} + 2\mh \tilde B_1 + \tilde \phi_1 - \Delta \tilde E_1)_{,k} + 2 \tilde E_{1 ,ikl}\tilde E_{1 ,jkl} -\\
    && 16\pi G \tilde \dvp_{1 ,i}\tilde \dvp_{1 ,j}].
\label{eeij:ug2}
\eea
Again, the right hand side of the above equation is identical to what we obtain in GR, apart from the fact that $E_1\neq 0$ in UG. In compact notation, the above equation can be expressed as 
\be
(\tilde B_2^{\prime} -\tilde E_2^{\prime\prime}) + 2\mh ( \tilde B_2-\tilde E_2^{\prime}) + \tilde \phi_2 = \partial_{i}^{-1}\partial_{j}^{-1} Y_{2 i j},
\label{eeij:ug21}
\ee
where $Y_{2 i j}$ is the terms inside the square bracket on the right hand side of (\ref{eeij:ug2}),
\bea
\nn Y_{2 i j} &=& 2(\tilde B_1-\tilde E_1^{\prime})_{,i j} \{\tilde \phi_1^{\prime} + \Delta(\tilde B_1-\tilde E_1^{\prime})\} + 2\tilde \phi_{1 ,i}\tilde \phi_{1 ,j} - 
 2 ( \tilde B_1-\tilde E_1^{\prime})_{, ik}( \tilde B_1-\tilde E_1^{\prime})_{, jk} - \\
\nn && 2 \tilde E_{1 ,ik}^{\prime}\tilde E_{1 ,jk}^{\prime} + 2\tilde E_{1 ,ijk}(\tilde B_1^{\prime} + 2\mh \tilde B_1 + \tilde \phi_1 - \Delta \tilde E_1)_{,k} + 2 \tilde E_{1 ,ikl}\tilde E_{1 ,jkl} -\\
&& 16\pi G \tilde \dvp_{1 ,i}\tilde \dvp_{1 ,j}.
\label{Y2}
\eea
It can be noticed that $Y_{2 i j}$ is a quadratic function of the first order quantities $\tb_1, \te_1, \tp_1, \tdvp_1$ and their spatial derivatives. Therefore, $Y_{2 i j}$ can be expressed in terms of the first order field fluctuation $\tdvp_1$ and its derivatives.

Finally, using (\ref{ugc1:uch}), the second order unimodular constraint equation (\ref{ugc:s}) on the uniform curvature hypersurfaces can be expressed as 
\be
\tilde\phi_2 + \Delta E_2 + 2 \tilde\phi_1^2 - (\tilde B_{1 ,i}\tilde B_{1 ,i}) - 2\tilde\phi_{1 ,ij}\tilde\phi_{1 ,ij} = 0.
\label{ugc:s:uch}
\ee

Using the background and the first order equations, the ($\eta, \eta$) component of the perturbed field equation can be written as 
\bea
\nn && 3\mh\tp_2^{\prime} + \Delta\{\tp_2 - \mh(\tb_2-\te_2^{\prime}) + \tb_2^{\prime} - \te_2^{\prime\prime}\} - 12\mh\tp_1\tp_1^{\prime} - 2\tp_1^{\prime} \Delta(\tb_1 - \te_1^{\prime}) + \\
\nn && 2\tp_{1 ,i}(\mh\tb_1 - \Delta\te_1)_{,i} - 2\tp_{1 ,i}\tp_{1 ,i} - 4\tp_{1 ,ij}\te_{1 ,ij} - 4E_{1 ,ij}(\tb_1^{\prime} - \te_1^{\prime\prime})_{,ij} + 2(\mh\tb_1 - \tb_1^{\prime})_{,i}\Delta\te_{1 ,i} + \\
\nn && 4\mh(\tb_1 - \te_1^{\prime})_{,ij}\te_{1 ,ij} - (\tb_1 - \te_1^{\prime})_{,ij}(\tb_1 - \te_1^{\prime})_{,ij} + 2 \te_{1,ij}^{\prime}\te_{1,ij}^{\prime} + \Delta(\tb_1 - \te_1^{\prime})\Delta(\tb_1 - \te_1^{\prime}) +\\
\nn && 6\mh\tb_{1 ,i}\tb_{1 ,i}^{\prime} - \Delta\te_{1 ,i}\Delta\te_{1 ,i} + \te_{1 ,ijk}\te_{1 ,ijk} = 8\pi G [3\vpb^{\prime}\tdvp_2^{\prime} + 6(\tdvp_1^{\prime 2}) + \tdvp_{1 ,i}\tdvp_{1 ,i} - \\ 
&& 2\vpb^{\prime}\tb_{1 ,i}\tdvp_{1 ,i}].
\label{eeee:ug2}
\eea
The above equation can be written in a compact form, where all the terms involving spatial derivatives of the first order quantities are separated included in the $Z_2$ quantity,
\be
3\mh\tp_2^{\prime} + \Delta\lsb\tp_2 - \mh(\tb_2-\te_2^{\prime}) + \tb_2^{\prime} - \te_2^{\prime\prime}\rsb - 12\mh\tp_1\tp_1^{\prime} = 8\pi G \lsb 3\vpb^{\prime}\tdvp_2^{\prime} + 6\lsp\tdvp_1^{\prime 2}\rsp\rsb + Z_2,
\label{eeee:ug21}
\ee
where $Z_2$ is given by
\bea
\nn Z_2 &=& 2\tp_1^{\prime} \Delta\lsp\tb_1 - \te_1^{\prime}\rsp - 2\tp_{1 ,i}\lsp\mh\tb_1 - \Delta\te_1\rsp_{,i} + 2\tp_{1 ,i}\tp_{1 ,i} + 4\tp_{1 ,ij}\te_{1 ,ij} - 2 \te_{1,ij}^{\prime}\te_{1,ij}^{\prime} -\\
\nn && 2\lsp\mh\tb_1 - \tb_1^{\prime}\rsp_{,i}\Delta\te_{1 ,i} - 4\mh\lsp\tb_1 - \te_1^{\prime}\rsp_{,ij}\te_{1 ,ij} + \lsp\tb_1 - \te_1^{\prime}\rsp_{,ij}\lsp\tb_1 - \te_1^{\prime}\rsp_{,ij} - \\
\nn && \Delta\lsp\tb_1 - \te_1^{\prime}\rsp\Delta\lsp\tb_1 - \te_1^{\prime}\rsp - 6\mh\tb_{1 ,i}\tb_{1 ,i}^{\prime} + \Delta\te_{1 ,i}\Delta\te_{1 ,i} - \te_{1 ,ijk}\te_{1 ,ijk} +\\
    && 4E_{1 ,ij}\lsp\tb_1^{\prime} - \te_1^{\prime\prime}\rsp_{,ij} + 8\pi G \lsb\tdvp_{1 ,i}\tdvp_{1 ,i} - 2\vpb^{\prime}\tb_{1 ,i}\tdvp_{1 ,i}\rsb.
\label{Z2}
\eea
Using (\ref{eeij:ug21}), equation (\ref{eeee:ug21}) can be simplified as
\be
\tp_2^{\prime} - \Delta\lsp\tb_2-\te_2^{\prime}\rsp = \frac{8\pi G}{\mh} \lsb\vpb^{\prime}\tdvp_2^{\prime} + 2\lsp\tdvp_1^{\prime 2}\rsp\rsb + 4\tp_1\tp_1^{\prime} +\frac{1}{3\mh} Z_2 - \frac{1}{3\mh}\Delta\lsp\partial^{-1}_i\partial^{-1}_j Y_{2ij}\rsp.
\label{eeee:ug22}
\ee
Again, one can see that this equation is identical in GR, apart from the fact that in UG, $\te_1$ is given by (\ref{ugc1:uch}).

We will use equations (\ref{eeei:ug2}), (\ref{eeij:ug21}) and (\ref{eeee:ug22}) to construct the equation of motion for the second order gauge invariant Mukhanov-Sasaki variable.

\subsection{Second order Gauge invariant Mukhanov-Sasaki variable}

Like in first order perturbation, one can express the transformations of energy density and field fluctuations in second order by using (\ref{tt:gen1}) and (\ref{lie:svt}),
\bea
\nn \tilde{\delta\rho_2}&=&\delta\rho_2 - \alpha_{2}\rho^{\prime}_{0} + \alpha_{1}\lsp\alpha_{1}\rho^{\prime \prime}_{0} + \alpha^{\prime}_{1}\rho^{\prime}_{0} - 2\delta\rho_1^{\prime}\rsp + \beta_{1 ,i}\lsp\alpha_{1}\rho^{\prime}_{0}-2\delta\rho_1\rsp_{,i},\\
    \tilde{\delta\vp_2}&=&\delta\vp_2 - \alpha_{2}\vp^{\prime}_{0} + \alpha_{1}\lsp\alpha_{1}\vp^{\prime \prime}_{0} + \alpha^{\prime}_{1}\vp^{\prime}_{0} - 2\delta\vp_1^{\prime}\rsp + \beta_{1,i}\lsp\alpha_{1}\vp^{\prime}_{0}-2\delta\vp_1\rsp_{,i}.
\label{gt:rph2}
\eea
The second order curvature and Newtonian potential transform respectively as 
\bea
\nn \tilde{\psi_2}&=&\psi_2 + \mh\alpha_{2} - \alpha_{1}\lsb\mh\alpha_{1}^{\prime} + \lsp\mh^{\prime}+2\mh^{2}\rsp\alpha_1 + 2\psi_1^{\prime} + 4\mh\psi_1\rsb - \beta_{1 ,i}\lsp\mh\alpha_{1}+2\psi_1\rsp_{,i},\\
\nn \tilde{\phi_2}&=&\phi_2 - \mh\alpha_{2} - \alpha_{2}^{\prime} + \alpha_{1}\lsb\alpha_{1}^{\prime\prime} + 5\mh\alpha_{1}^{\prime} + \lsp\mh^{\prime}+2\mh^{2}\rsp\alpha_{1} - 4\mh\phi_1 -2\phi_1^{\prime}\rsb + \\
    && \alpha_{1}^{\prime}\lsp 2\alpha_{1}^{\prime} - 4\phi_1\rsp + \beta_{1 ,i}\lsp\alpha_{1}^{\prime} + \mh\alpha_{1} - 2\phi_1\rsp_{,i} + \beta_{1 ,i}^{\prime}\lsp\alpha_{1} + 2B_1 - \beta_{1}^{\prime}\rsp_{,i}.
\label{gt:phsi2}
\eea

On uniform curvature hypersurfaces (where $\tilde{\psi_2}=\tilde{\psi_1}=0$), we use (\ref{gt:phsi2}) and we rewrite $\alpha_2$ as 
\be
\alpha_2=-\frac{\psi_2}{\mh} + \frac{\alpha_{1}}{\mh}\lsb\mh\alpha_{1}^{\prime} + \lsp\mh^{\prime}+2\mh^{2}\rsp\alpha_1 + 2\psi1^{\prime} + 4\mh\psi1\rsb + \frac{1}{\mh}\beta_{1 ,i}\lsp\mh\alpha_{1}+2\psi1\rsp_{,i},
\label{al2:uch}
\ee
where $\alpha_1$ is given by (\ref{al1:uch}). We consequently get
\be
\alpha_2 = - \frac{\psi_2}{\mh} - \lsb\frac{\psi_1}{\mh^2}\lsp\psi_1^{\prime}+2\mh\psi_1\rsp\rsb + \frac{1}{\mh}\beta_{1 ,i}\psi_{1, i}.
\label{al2:uch1}
\ee
Substituting $\alpha_1$ and $\alpha_2$ in (\ref{gt:rph2}), the second order Mukhanov-Sasaki variable ($\mathcal{Q}_2$) can be expressed as
\bea
\nn \mathcal{Q}_2 = \tdvp_{2}|_{\tilde{\psi_2}=0} &=& \lsp\dvp_2 + \frac{\vp^{\prime}_0}{\mh}\psi_2\rsp + \lsp\frac{\psi_1}{\mh}\rsp^2\lsp\vp_0^{\prime\prime}+2\mh\vp_0^{\prime}-\vp_0^{\prime}\frac{\mh^{\prime}}{\mh} \rsp + 2\frac{\vp_0^{\prime}}{\mh^2}\psi_1\psi_1^{\prime} +\\
&& 2\frac{\psi_1}{\mh}\dvp_1^{\prime} - 2\beta_{1 ,i}\lsp \dvp_1+\frac{\vp_0^{\prime}}{\mh}\psi_1\rsp_{, i}.
\label{ms2:uch}
\eea

Compared to first order, second order Mukhanov-Sasaki variable is complicated because of the presence of the spatial part of first order coordinate transformation $\beta_1$. In this context we make the following comments:
\begin{itemize}
\item It should be noticed that $\tdvp_2$ in Eq. (\ref{gt:rph2}) only depends upon the gauge generators, $\alpha_2$, $\alpha_1$ and $\beta_1$. The expression of $\tdvp_2$ is identical in both UG and GR.

\item 
In UG, $\alpha_1$ is fixed in the same way as in GR, \textit{i.e.} $\alpha_1=-\frac{\psi_1}{\mh}$. As a consequence, the expression of $\alpha_2$, given in (\ref{al2:uch1}) is also identical in UG and GR. Therefore, the expression of $\mathcal{Q}_2$ given in (\ref{ms2:uch}) is identical in UG and GR.
\item In GR, $\beta_1$ is fixed by setting $\tilde{E_1} = 0$, \textit{i.e.} $\beta_1=E_1$. In contrast, in UG, $\beta_1$ is fixed by first order unimodular constraint (\ref{ugc:f1}), \textit{i.e.} $\beta_1=-\Delta^{-1}\lsp\alpha_1^{\prime}+4\mh\alpha_1\rsp$. 
\item However, as discussed earlier, it is to be noted that in GR, $\beta_1=E_1$ or $\tilde{E_1} = 0$ is a choice. Therefore, even in GR, the definitions of $\mathcal{Q}_2$ can be different depending upon the different choices of $\beta_1$ or $\te_1$. In GR, if one chooses $\beta_1=-\Delta^{-1}\lsp\alpha_1^{\prime}+4\mh\alpha_1\rsp$, the expression of second order Mukhanov-Sasaki variable will be exacly same as in UG. 

\end{itemize}

As a consequence, in second order perturbation, only one choice of Mukhanov-Sasaki variable in GR will be same as UG. The other choices of second order Mukhanov-Sasaki variables will be different. It is interesting to note that, once $\alpha_2$ is fixed by (\ref{al2:uch1}), spatial part of second order coordinate transformation $\beta_2$ will be fixed from second order unimodular constraint. As $\alpha_2$ and $\tdvp_2$ are independent of $\beta_2$, second order unimodular constraint does not have any impact on $\mathcal{Q}_2$. Second order unimodular constraint will be used in the construction of Mukhanov-Sasaki variable higher than second order. In the rest of the Section we will show that this will also be true for the evolution equation of second order Mukhanov-Sasaki variable.  

\subsection{Evolution equation of second order Mukhanov-Sasaki variable}
Using the background (\ref{back:frd}) and first order (\ref{kg1:uch}) equations, the equation of motion of the second order field fluctuation $\tdvp_2$ on constant curvature hypersurfaces becomes,
\bea
\nn && \tdvp_2^{\prime\prime} + 2\mh\tdvp_2^{\prime} - \Delta\tdvp_2 + a^2V_{,\vp\vp}\tdvp_2 - \vp_0^{\prime}\tp_2^{\prime} + 2a^2V_{,\vp}\tp_2 - \vp_0^{\prime}\Delta\lsp\tb_2 - \te_2^{\prime}\rsp + a^2 V_{,\vp\vp\vp} \tdvp_1^2 - \\
\nn && 2\vp_0^{\prime}\tb_{1,i}\tb_{1,i}^{\prime} + 2a^2 V_{,\vp}\tb_{1,i}\tb_{1,i} - 4\tb_{1,i}\tdvp_{1,i}^{\prime} + 2\vp_0^{\prime}\tb_{1,i}\tp_{1,i} + 4\vp_0^{\prime}\lsp\tb_1 - \te_1^{\prime}\rsp_{,ij}\te_{1,ij} +  \\
\nn && 2\vp_0^{\prime}\tb_{1,i}\Delta\te_{1,i} + 2\tdvp_{1,i}\Delta\te_{1,i} + 4\tdvp_{1,ij}\te_{1,ij} - 4\mh\tb_{1,i}\tdvp_{1,i} - 2\tdvp_{1,i}\tb_{1,i}^{\prime} -  2\tdvp_{1,i}\tp_{1,i} -\\
 && 2\tp_1^{\prime}\tdvp_1^{\prime} - 2\tdvp_1^{\prime}\Delta\lsp\tb_1-\te_1^{\prime}\rsp + 4\vp_0^{\prime}\tp_1\tp_1^{\prime} - 4\tp_1\Delta\tdvp_1 + 4 a^2 V_{,\vp\vp}\tp_1\tdvp_1 = 0.
\label{kg2:ug}
\eea
In the above equation, $\tdvp_2$ can be replaced as $\mathcal{Q}_2$. Further, $\tp_2$, $\tb_2$ and $\te_2$ can be eliminated from the equation of motion by using equations (\ref{eeei:ug2}) and (\ref{eeee:ug22}). Finally, equation (\ref{kg2:ug}) can be expressed in terms of the second order Mukhanov-Sasaki variable ($\mathcal{Q}_2$) as 
\bea
\nn && \mathcal{Q}_2^{\prime\prime} + 2\mh\mathcal{Q}_2^{\prime} - \Delta\mathcal{Q}_2 + \lsb a^{2} V_{, \vp \vp} + 8\pi G \lsc a^{2} V_{,\vp}
\frac{\vpb^{\prime}}{\mh} - 
\vpb^{\prime}\lsp\frac{\vpb^{\prime}}{\mh}\rsp^{\prime}
\rsc 
\rsb\mathcal{Q}_2 + 2a^2 V_{,\vp}W_2 - 2\vpb^{\prime}W_2^{\prime} +\\
&& \frac{\vpb^{\prime}}{3\mh}Z_2 - \frac{\vpb^{\prime}}{3\mh}\Delta\lsp\partial_i^{-1}\partial_j^{-1}Y_{2ij}\rsp + U_2= 0,
 \label{kg2:ug1}
\eea
where
\bea
\nn W_2 &=& \lsp 8\pi G \frac{1}{\mh}\tdvp_1\tdvp_1^{\prime} + 2\tp_1^2 - \tb_{1,i}\tb_{1,i} + \frac{1}{\mh}\Delta^{-1}X_2\rsp,\\
\nn U_2 &=& 16\pi G\frac{\vpb^{\prime}}{\mh}\lsp\tdvp_1\rsp^2 + 4\vpb^{\prime}\tp_1\tp_1^{\prime} + a^2 V_{,\vp\vp\vp}\tdvp_1^2 - \\
\nn && 2\vp_0^{\prime}\tb_{1,i}\tb_{1,i}^{\prime} + 2a^2 V_{,\vp}\tb_{1,i}\tb_{1,i} - 4\tb_{1,i}\tdvp_{1,i}^{\prime} + 2\vp_0^{\prime}\tb_{1,i}\tp_{1,i} + 4\vp_0^{\prime}\lsp\tb_1 - \te_1^{\prime}\rsp_{,ij}\te_{1,ij} +  \\
\nn && 2\vp_0^{\prime}\tb_{1,i}\Delta\te_{1,i} + 2\tdvp_{1,i}\Delta\te_{1,i} + 4\tdvp_{1,ij}\te_{1,ij} - 4\mh\tb_{1,i}\tdvp_{1,i} - 2\tdvp_{1,i}\tb_{1,i}^{\prime} -  2\tdvp_{1,i}\tp_{1,i} -\\
 && 2\tp_1^{\prime}\tdvp_1^{\prime} - 2\tdvp_1^{\prime}\Delta\lsp\tb_1-\te_1^{\prime}\rsp + 4\vp_0^{\prime}\tp_1\tp_1^{\prime} - 4\tp_1\Delta\tdvp_1 + 4 a^2 V_{,\vp\vp}\tp_1\tdvp_1 .
\eea
In the above equation $X_2$, $Y_{2ij}$ and $Z_2$ are given by (\ref{X2}), (\ref{Y2}) and (\ref{Z2}) respectively. 

The equation of motion of the second order Mukhanov-Sasaki variable consists of two parts, one part containing $\mathcal{Q}_2$, and the other part containing quadratic functions of first order quantities --- ($\tb_1$), ($\te_1$) and ($\tdvp_1$). As $\tdvp_1=\mathcal{Q}_1$, using (\ref{feei:uch1}) and (\ref{tbte1:uch}), the perturbed second order Klein-Gordon equation (\ref{kg2:ug1}) can be expressed entirely in terms of $\mathcal{Q}_2$ and a quadratic function of $\mathcal{Q}_1$.

It should be noticed that, similarly to the first order case, in the differential equation of $\tdvp_2$ (\ref{kg2:ug}), the terms $\tb_2$ and $\te_2$ come as a linear combination of ($\tb_2 - \te_2^{\prime}$). As a consequence, the derivation of equation (\ref{kg2:ug1}) does not require the second order unimodular constraint given by (\ref{ugc:s:uch}). The second order unimodular constraint is only needed to calculate $\tb_2$ and $\te_2$ individually, which might be useful in third order or higher order perturbation theories. However, from the expressions of $U_2, W_2, X_2, Y_{2ij}$ and $Z_2$, it can be noticed that all the terms containing first order quantities $\tb_1$ and $\te_1$ can not be written as a linear combination ($\tb_1 - \te_1^{\prime}$). Therefore, the final equation of $\mathcal{Q}_2$ will be dependent on $\tb_1$ and $\te_1$ given in (\ref{tbte1:uch}), just like the quantity $\mathcal{Q}_2$ itself. As equation (\ref{kg2:ug1}) is also valid for GR, the evolution equation of $\mathcal{Q}_2$ will be same in GR and UG, provided that one only chooses $\te_1$ as given in (\ref{tbte1:uch}) in GR. Any other choices of $\te_1$ other than (\ref{tbte1:uch}) will lead to a discrepancy between GR and UG in second order. 

%%%%%%%%%%%%%%%%%%%%%%%%%%%%%%%%%%%%%%%%%%%%%%%%%%%%%%%%%%%%%%%%%%%%%%%%%%%%%%%%%%%%%%%%%%%%%%%%%%%%%
\section{Conclusion and discussion}\label{conclusion}

In the background trace-free part of the Einstein equations, the information of the potential appears through the energy conservation equation. Therefore, inflation can occur in UG in the same way as in GR. In UG, the cosmological constant comes as an integration constant. Hence, it can be fixed by observations.
In this work we have performed a detailed analysis of the first order and second order perturbation of UG. Using the unimodular constraints, various gauge invariant quantities have been calculated in first and second order perturbation and the results are compared with GR. 

In first order perturbation it has been shown that, the gauge invariant definitions of metric perturbations $\Phi_1$ and $\Psi_1$ are same as in GR. Only difference between UG and GR is the choices of the hypersurfaces. In GR, to construct $\Phi_1$ and $\Psi_1$ we need choose hypersurfaces such that $\tb_1=\te_1=0$. On the other hand, in UG, because of the unimodular constraint, the same $\Phi_1$ and $\Psi_1$ quantities can be constructed on $\tb_1=\te_1^{\prime}$ hypersurfaces. In first order perturbation, we have also shown that unimodular constraint does not affect the other gauge invariant quantities such as $\mathcal{R}_1$, $\mathcal{Q}_1$, $\zeta_1$ and $\mathcal{M}_1$. In this work, it is shown that, as first order unimodular constraint does not have any impact, the evolution equation of first order Mukhanov-Sasaki variable $\mathcal{Q}_1$ is exactly same as in GR. The evolution of the gauge invariant quantity $\zeta_1$ is identical in GR and UG. 

Because of the divergence-less and trace-less conditions, the vector and tensor modes are not affected by the unimodular constraint. Hence, the evolution of vector and tensor modes are also identical in UG and GR. 

We have calculated the first order contribution in CMB temperature anisotropies on large scales in UG. In contrast to some of the earlier work\cite{Gao:2014nia}, we do not find any dipole term in the temperature fluctuations due to UG. The expression of CMB temperature fluctuation in UG on large scales is given as $\frac{\delta T_1}{T_0}=\frac{1}{3}\Phi_1$, which is exactly same as obtained in GR. This result is also supported by \cite{peter2009primordial}, where CMB temperature anisotropies are calculated in generic gauge. 

In second order perturbation, Mukhanov-Sasaki variable ($\mathcal{Q}_2$) is calculated in UG. Here we have shown that, in second order perturbation $\mathcal{Q}_2$ and its evolution equation are not dependent upon the second order unimodular constraint but, the first order unimodular constraint. Although, $\mathcal{Q}_2$ and its evolution are affected by the first order unimoular constraint, this does not imply any significant difference between the two theories. In GR, setting $\te_1=0$ is a choice. If in GR, $\tb_1$ and $\te_1$ are set to the expressions given in (\ref{tbte1:uch}), $\mathcal{Q}_2$ and its evolution becomes exactly same as derived in UG. This only means sacrificing flat slicing gauge in GR \textit{i.e.}, second order 3-Ricci scalar $^{(3)}R_2\neq 0$.

Therefore, in the light of our analyses of first order and second order perturbations, it can be concluded that GR and UG are equivalent. The only difference lies in the gauge choices. In UG, we have restricted gauge choices compared to GR. However, this does not affect the observations as we always relate the observables with the gauge invariant quantities. The same conclusions will still be true for perturbation at a higher order than second order. 

%%%%%%%%%%%%%%%%%%%%%%%%%%%%%%%%%%%%%%%%%%%%%%%%%%%%%%%%%%%%%%%%%%%%%%%%%%%%%%%%%%%%%%%%%%%%%%%%%%%%%
\section{Acknowledgements}
We thank Kasper Peeters for computer algebra system Cadabra \cite{peeters2007introducing,peeters2006field} which was useful for verifying our calculations.
We thank Swastik Bhattacharya and Debottam Nandi for useful discussions. 
The work is supported by Max Planck-India Partner Group on Gravity and Cosmology. SS is supported by Ramanujan Fellowship of DST, India.

%%%%%%%%%%%%%%%%%%%%%%%%%%%%%%%%%%%%%%%%%%%%%%%%%%%%%%%%%%%%%%%%%%%%%%%%%%%%%%%%%%%%%%%%%%%%%%%%%%%%%
\bibliography{TFEE}
\bibliographystyle{JHEP}

\end{document}